\documentclass[12pt,a4paper,notitlepage,amsmath]{iopart}
\usepackage{myaasmacros}
\usepackage{lscape}
\usepackage{multirow}
\usepackage{iopams}
\usepackage{tensor}
\usepackage{graphicx}
\usepackage{bm}

\newcommand{\matrthree}[6]{{
{\tiny $\left( \begin{array}{ccc} \! #1 \! & \! #2 \! & \! #3 \! \\
 \! #2 \! & \! #4 \! & \! #5 \! \\
\! #3 \! & \! #5 \! & \! #6 \! 
\end{array}\right)$}}}

\newcommand{\matr}[1]{{ \left(\begin{array}{ccc} \! #1 \! \end{array} \right) }}
\newcommand{\vecthree}[3]{{
{\tiny $\left( \begin{array}{c} \! #1 \! \\
\! #2 \! \\
\! #3 \!
\end{array}\right)$}}}

\renewcommand{\vec}{\bi}

\begin{document}

\newcommand{\dd}{\mathrm{d}}
\newcommand{\udot}{\dot{\hspace{2mm}}}
\newcommand{\lie}{\mathcal{L}}
\newcommand{\pricomm}[1]{}
\newcommand{\ispace}{\phantom{\hat{b}}\hspace{-1mm}}

\title{Linearization of homogeneous, nearly-isotropic cosmological models} \author{Andrew
  Pontzen$^1$, Anthony Challinor$^{1,2}$} \address{$^1$ Institute of
  Astronomy and Kavli Institute for Cosmology Cambridge,
  Madingley Road, Cambridge CB3 0HA, UK} \address{$^2$
  DAMTP, Centre for Mathematical Sciences, Wilberforce Road, Cambridge
  CB3 0WA, UK } \ead{apontzen@ast.cam.ac.uk}

\begin{abstract}
  Homogeneous, nearly-isotropic Bianchi cosmological models are
  considered.  Their time evolution is expressed as a complete set of
  non-interacting linear modes on top of a Friedmann-Robertson-Walker
  background model.  This connects the extensive literature on Bianchi
  models with the more commonly-adopted perturbation approach to
  general relativistic cosmological evolution.  Expressions for the
  relevant metric perturbations in familiar coordinate systems can be
  extracted straightforwardly.  Amongst other possibilities, this
  allows for future analysis of anisotropic matter sources in a more
  general geometry than usually attempted.

 We discuss the geometric mechanisms by which maximal symmetry is
 broken in the context of these models, shedding light on the origin
 of different Bianchi types. When all relevant length-scales are
 super-horizon, the simplest Bianchi I models emerge (in which
 anisotropic quantities appear parallel transported).

 Finally we highlight the existence of arbitrarily long
 near-isotropic epochs in models of general Bianchi type (including
 those without an exact isotropic limit).

\end{abstract}
\maketitle

\section{Introduction}

Mathematically, the simplest models of the Universe are those which
possess a high degree of symmetry. Spacelike slices of the
Friedmann-Robertson-Walker (FRW) solutions possess the maximal six
Killing vector fields (KVFs), reducing the number of degrees of
freedom to just one, the curvature radius. A historical method of
relaxing this assumption to study more complex models is to remove
three KVFs corresponding to rotational invariance, leaving only a
single group which acts simply transitively on the spacelike
sections. This yields the set of ``Bianchi'' models, named after the
classification scheme for three-parameter Lie groups
\cite{bianchi1897}. One of the advantages of this approach is that the
evolution (Einstein) equations reduce to ordinary differential
equations.  This means that the full, non-linear behaviour of Bianchi
universes is open to study by analytic methods (e.g. dynamical systems
approaches \cite{wainwright1997dsc}) or numerical integration.

Observational interest in these anisotropic models has receded
somewhat since the publication of full-sky maps of the temperature
anisotropies in the cosmic microwave background (CMB) by COBE
\cite{1992ApJ...396L...1S} and WMAP
\cite{2003ApJS..148....1B,2010arXiv1001.4744J}. It is known
\cite{1995ApJ...443....1S} that minimal assumptions\footnote{It is
  worth noting that these assumptions include a temporal Copernican
  assumption; arbitrary large-scale anisotropies may be concealed from
  observation on a single time-slice.}  coupled with the observed
near-isotropy of the CMB suggest that our universe has been highly
isotropic since the last scattering surface at a redshift of $z \sim
1100$. Of course, this does not preclude the existence of mild
anisotropy at a level below (or bordering on) our current detection
threshold.

In fact known `anomalies' in the CMB data have been shown to be
mimicked by a weak signal corresponding to the pattern expected in
VII$_h$ universes
\cite{2005ApJ...629L...1J,2006ApJ...644..701J,2007MNRAS.377.1473B}.
Such models are, in the consensus view, unrealistic in that they
require an unfeasibly large value of $\Omega_K$ in light of strong
constraints from other observations including the Gaussian random
fluctuations in the CMB itself \cite{2007ApJS..170..335P}. Further,
the expected polarisation signals from a Bianchi component are
apparently incompatible with the limits on $B$-mode signals and
$TB$-correlations in WMAP \cite{2007MNRAS.380.1387P}. On the other
hand, the dynamical model employed in these studies ignored certain
degrees of freedom in the interest of simplicity. One result is that
the shear principal axes have a fixed alignment relative to the
residual symmetry axis of the models -- this restricts the appearance
of the Bianchi component of the CMB to a form which is not entirely
generic
\cite{2007MNRAS.380.1387P,2009PhRvD..79j3518P,2010arXiv1004.0957S}.
In a full analysis, one further finds that certain sub-models have
oscillatory dynamics, rather than the monotonically decaying shear
seen in restricted solutions.  This certainly permits the polarisation
constraint to be evaded \cite{2009PhRvD..79j3518P} and may allow for a
self-consistent anisotropic Bianchi universe to provide an improved
fit to the CMB data over concordance models. However, a proper
statistical analysis to support such an optimistic statement is
currently lacking, and it is more plausible that, even with the
additional freedoms, one will not find a model consistent with all
known constraints. 

On the theoretical side, the Bianchi models remain useful pedagogical
tools regardless of their observational status.  Furthermore the
emergence of anisotropic pocket universes in string-inspired eternal
inflation models \cite{2010JCAP...07..007B} reminds us that the
simple, globally isotropic picture of the present-day cosmos may yet
be replaced.

The work described here is intended to clarify the status of Bianchi
models which are close to isotropy. Modern approaches to observational
cosmology generally impose linear perturbations on a fully isotropic
(FRW) background.  Bianchi models which are almost isotropic must be
expressible in this way: as small, `homogeneous' (in a sense to be
defined) perturbations on an appropriate FRW background.  Such
decompositions of Bianchi models into non-interacting modes has been
made for individual situations in the past, and in some isolated cases
can be made exact~(e.g. Refs
\cite{1973ApJ...180..317C,1975JETP...42..943G,1976NCimB..35..268L,1986PhR...139....1B,1991PhRvD..44.2356K});
however, here we exhibit a method for understanding {\it all}
available homogeneous linear perturbations about an exactly isotropic
background.  The mode decomposition we present is very naturally
suited to use in observational studies and has already been combined
with the CMB transport equations in Ref. \cite{2007MNRAS.380.1387P} to
calculate consistently the full range of CMB patterns in
nearly-isotropic Bianchi universes~\cite{2009PhRvD..79j3518P}.

The rest of this work is structured as follows. In Section
\ref{sec:what-homogeneity} we discuss the origin of the Bianchi model
representations of maximally-symmetric spaces, highlighting how
ambiguities arise in the notion of translations and the particular
difficulty in resolving these ambiguities in an open space. We expand
these ideas in concrete terms for each space in Section
\ref{sec:coord-expr-modes}, with some further detail on alternative
coordinate systems in \ref{sec:altern-coord-expr}.  In Section
\ref{sec:enum-line-bianchi}, we perform the decomposition of suitable
Bianchi models into a background with linearised modes and discuss the
classification of such modes alongside their relation to the
automorphism formalism. We relate our results to previous work in
Section \ref{sec:stab-line-relat}. Finally, in Section
\ref{sec:other-bianchi-types}, we outline the properties of a class of
almost-isotropic Bianchi models which do not possess the fiducial
isotropic limits. We provide a brief summary in Section
\ref{sec:summary}.

We use a metric signature $-+++$ throughout. Latin indices represent
spatial components ($1 \to 3$) while Greek indices represent spacetime
components ($0 \to 4$). Indices relative to an orthonormal frame are
given a hat ($\hat{a}, \hat{b}, \dots$). Units are adopted so that
$c=1=8\pi G$ throughout.

The reader who wishes to skip the derivation and discussion will find
a concise list of Bianchi perturbations in Table
\ref{tab:frw-perturb}, their evolution and constraints described by
equations~(\ref{eq:eigenmode-evol}) and (\ref{eq:eigenmode-tilt}), and
coordinate expressions to expand such modes relative to a given
coordinate system in Section \ref{sec:coord-expr-modes}.

\section{Homogeneous, anisotropic models}\label{sec:what-homogeneity}

\subsection{Notational background; the traditional approach}

Given a manifold $\mathcal{M}$ with metric $g$, symmetries are
mathematically expressed by the invariance of $g$ under appropriate
transformations of $\mathcal{M}$.  Continuous transformations form a
Lie group of operations on $\mathcal{M}$; it is consequently possible
to show that all continuous group operations are built from
infinitesimal generators, and that invariance under the
infinitesimal generators is sufficient to prove invariance under the
full group. Invariance of the metric $g$ under infinitesimal
transformations is thus encoded in the notion of a vanishing Lie
derivative along the tangent vector field:
\begin{equation}
\mathcal{L}_{\mathbf{{\bm \xi}}} g = 0 \label{eq:kvf}
\end{equation}
where $\mathbf{{\bm \xi}}$ is a vector field everywhere tangent to the
motion, expressing the movement of each point towards a neighbour.  A
vector field satisfying condition (\ref{eq:kvf}) is known as a Killing
vector field (KVF). The linearity of the Lie derivative means that
linear combinations of KVFs are Killing. Furthermore, the Jacobi
identities can be used to show that the commutator of two KVFs is also
Killing, {\it i.e.} $\mathcal{L}_{[\mathbf{{\bm \xi}}_1,\mathbf{{\bm \xi}}_2]} g=0$.

In the context of Bianchi models, homogeneity is defined by the
existence of a closed algebra of three Killing vector fields
which are everywhere linearly independent (so that the isometry group
is said to act simply transitively on $\mathcal{M}$).  Intuitively,
this expresses the existence of paths covering the entire space, along
which one may walk from any point to any other while seeing no change
in the metric. The closure of the algebra is expressed by demanding
that
\begin{eqnarray}
  \left[{\bm \xi}_i,{\bm \xi}_j\right] = -C_{ij}^k {\bm \xi}_k \, ,\label{eq:kvf-closure}
\end{eqnarray}
where the $C_{ij}^k$ are constants and we have introduced a minus sign
for later convenience.  For $n=3$ spatial dimensions (which we will
assume throughout), it is conventional to decompose the structure
constants $C_{ij}^k$ as
\begin{equation}
C_{ij}^k = -a_i \delta_j^k + a_j \delta_i^k + \epsilon_{ijl} n^{lk}\textrm{,}
\end{equation}
where $n^{ij}$ is symmetric.  By reparametrizing the ${\bm \xi}_i$ with a
linear transformation of the form ${\bm \xi}_i \to \gamma_i^j {\bm \xi}_j$, one
may simultaneously require $n=\mathrm{diag}(n_1,n_2,n_3)$ and set
$\vec{a}=(a,0,0)$; in this scheme, the Jacobi identities reduce to
$an_1=0$. The standard approach is to partition the space of allowed
$n_1, n_2, n_3$ and $a$ into distinct regions or `Bianchi types';
models within each type are not necessarily isomorphic spaces but (up
to parity) they are deformable into each other by continuous
transformations of the metric at a point while keeping the Killing
fields fixed.  Conversely, models falling into different types are not
related by such a continuous transformation; it follows that time
evolution preserves the Bianchi type. We may further determine which
of the types permit isotropic sub-cases, and these are listed in Table
\ref{tab:frw-bianchi}. (See, for a concise derivation,
Ref. \cite{2007MNRAS.380.1387P}.)

\begin{table}
\begin{center}
\begin{tabular}{c|cccc|l}
Type & $a$ & $n_1$ & $n_2$ & $n_3$ & $\tensor[^{3}]{R}{_{\mathrm{FRW}}}\,e^{2\alpha}$ \\
\hline
I & 0 & 0 & 0 & 0 & 0 \\
V & $1$ & 0 & 0 & 0 & $-6$ \\
VII$_0$ & 0 & 0 & 1 & 1 & 0 \\
VII$_h$ & $\sqrt{h}$ & 0 & 1 & 1 & $-6h$ \\
IX & 0 & 1 & 1 & 1 & $3/2$ \\
\end{tabular}
\caption{Bianchi groups with FRW limit and their structure constants
  in canonical form with right-handed parity for the KVFs.  The final
  column is the comoving 3-curvature scalar in the $\beta=0$ FRW
  limit.}\label{tab:frw-bianchi}
\end{center}
\end{table}

\subsection{An alternative approach: explicitly breaking maximal symmetry}

The explanations given above form the standard approach to defining
the Bianchi spaces. However, for near-isotropic cases, it can be unclear
how the symmetries thus derived relate to more familiar symmetries of
the exactly-isotropic FRW cases: why does a given maximally-symmetric
model arise within certain Bianchi classes and not others?

To answer this question, it is instructive instead to \emph{start} with
the isometry group of the maximally-symmetric cases, and then inspect
the origin of the Bianchi-homogeneity and isotropy
subgroups. This gives an alternative view of constructing
near-isotropic Bianchi models: namely, we start with a
maximally-symmetric 3-space, identify the three Killing fields defining
homogeneity, and finally add a metric perturbation which is invariant
under the action of these fields (but not under the remainder of the
original isometry group). Dependent on the particular properties of
these preferred Killing fields, the perturbed space then falls 
into a specific Bianchi class. (Alternatively, one can
introduce perturbations which are invariant only under the isotropy
subgroup, which would lead to a Lema\^{i}tre-Tolman-Bondi solution.)

For a maximally-symmetric 3-space, a counting exercise shows there to
be six KVFs of which precisely three must vanish at any single chosen
point $p$. Since tracing the integral curves of these
locally-vanishing KVFs leaves $p$ fixed, the corresponding algebra
elements must generate the isotropy group of that point. We can
therefore immediately split the KVFs into two sets of three,
$\{\mathbf{R}_i\}$ (rotations) and $\{\mathbf{T}_i\}$ (translations),
where $\left. \mathbf{R}_i \right|_p=0$ and $\left. \mathbf{T}_i
\right|_p\ne 0$. Up to reparametrizations, the set of isotropy vector
fields for $p$ are uniquely defined. In fact, one can fix most of the
reparametrization freedom by demanding (without loss of generality)
that $\mathbf{R}_i$ satisfies the canonical commutation relations for
rotation generators,
\begin{equation}
[\mathbf{R}_i, \mathbf{R}_j] = - \epsilon_{ijk} \mathbf{R}_k \textrm{.}
\end{equation}
On the other hand the translations $\{\mathbf{T}_i\}$ have much more
freedom. While we can fix part of the reparametrization freedom by
demanding (again without loss of generality) that the $\mathbf{T}_i$
form an orthonormal basis at $p$, there remains the freedom
\begin{equation}
\mathbf{T}_i \to \mathbf{{\bm \xi}}_i = \mathbf{T}_i + \rho_i^j \mathbf{R}_j \textrm{,}\label{eq:rot-coeffs}
\end{equation}
where $\rho_i^j$ are constants.  It is this freedom which ultimately
leads to certain maximally-symmetric spaces falling into multiple
Bianchi types (Table \ref{tab:frw-bianchi}): there is no natural way
to pick out a preferred set of generators
$\mathbf{{\bm \xi}}_i$.\footnote{From a group-theoretic standpoint, the
  manifold is identified with the cosets of the rotation group; but
  because the rotations do not form a normal subgroup, there is no
  natural composition law for the cosets.}

Not all choices of $\rho_i^j$ are permitted: the three vector fields
${\bm \xi}_i$ must be closed [expression (\ref{eq:kvf-closure})] and also
everywhere linearly independent. If the former condition is not
obeyed, self-consistent ({\it i.e.}\ path-independent) transport will only be
available for tensors which are at least partially rotationally
symmetric, which is not our intention here (and leads either to full
isotropy or the intermediate Kantowski-Sachs models). If the latter
condition fails, the subgroup generated will not act transitively,
which also implies rotational symmetry about some point. One may get
further with determining the allowed commutation relations in a quite
general framework, but in fact it is more enlightening to adopt
explicit flat, open and closed models. These three types exhaust the
possibilities; this can be seen, for instance, by considering the need
for constant scalar curvature in maximally-symmetric spaces. We
perform the explicit search for homogeneity groups in Section
\ref{sec:coord-expr-modes}, finding exactly the groups of Table
\ref{tab:frw-bianchi} from this alternative approach.

Recalling that any tensor $x$ will be defined as homogeneous if it is
invariant under the action of the translational subgroup
($\mathcal{L}_{\vec{{\bm \xi}}_i} x=0$), we can now ask how such a
mathematical description compares with an intuitive notion of
homogeneity.  In flat space, for instance, physicists would almost
automatically assume that homogeneity implies invariance under the
Euclidean translational group (leading to a Type-I anisotropic
model). The most natural way to extend this to curved spacetimes is to
require our defining translational group to be generated by vector
fields which are tangent to a congruence of geodesics. By enumeration,
we will see that only Type-I (flat) and Type-IX (closed) models can
have this property.

Even where this can be arranged, anisotropic quantities on curved
backgrounds ({\it i.e.} in the Type-IX case) cannot simply turn out to be
parallel transported, since the latter arrangement is
path-dependent. Nonetheless, there are two special properties of Type-IX and Type-I symmetries of maximally-symmetric models which make them attractive, as we now outline.

First, for a Killing field to be geodesic, it is necessary and
sufficient that it have constant norm {\it i.e.} $\partial_{\alpha}
(|{\bm \xi}_i|^2) = 0$.  Hence the translations described by such a field
move each point by a constant geodesic distance and are known as
Clifford translations. This property can be used to demonstrate the
impossibility of constructing such Killing fields on an open
background \cite{wolf1972gac}.

Second, the group of translations in the Type-I and IX cases are
closed under rotation (the groups are normal within the full isometry
group, with an ideal subalgebra: $[\mathbf{R}_i,{\bm \xi}_j] =
-\epsilon_{ijk} {\bm \xi}_k$). In fact, a set of transitive Killing fields
on a maximally-symmetric space which are closed under rotation in this
way necessarily describe a congruence of geodesics. The converse can
also be shown to be true, {\it i.e.} a transitive set of Killing
vector fields is necessarily closed under rotations if each field is
tangent to geodesic congruences.\footnote{%
  On a maximally-symmetric 3-space with (constant) Ricci curvature
  $R$, one can show that $\nabla_a {\bm \xi}_b = \pm \sqrt{R/6}\,
  \eta_{abc}{\bm \xi}^c$ for a Killing field everywhere tangent to a
  congruence of geodesics. (Here $\eta_{abc}$ is the alternating
  tensor.) Such a Killing field is therefore determined solely by its
  value at some point which we take to be $p$ where the $\mathbf{R}_i$
  vanish. By isotropy of the space, the Killing field that results
  from action of the $\mathbf{R}_i$ on ${\bm \xi}_j$ is therefore equivalent
  to the field generated from the rotated version of ${\bm \xi}_j$ at $p$,
  {\it i.e.}\ a linear combination of the ${\bm \xi}_k$. This establishes closure
  under rotations.  } An observational consequence is that the CMB in
Type-I and Type-IX models has only dipolar or quadrupolar temperature
anisotropies\footnote{For the first order CMB the geodesics can be
  taken from the background FRW spacetime
  \cite{1985MNRAS.213..917B}. This being conformally related to
  $\mathbb{R} \otimes \mathcal{M}$, the photon null geodesics follow
  the pattern of spatial geodesics on
  $\mathcal{M}$.}~\cite{1985MNRAS.213..917B,2007MNRAS.380.1387P}.  (An
exception to this statement arising in a locally isotropic limit
falling outside the framework expounded here is explained in Section
\ref{sec:other-bianchi-types}.)

Conversely complex patterns in the CMB arise in Types V, VII$_0$ and
VII$_h$ precisely from the lack of global rotational symmetry in the
propagation equations for anisotropic quantities. We can now understand
this to be a consequence of any of three distinct but equivalent statements:
\begin{itemize}
\item the Killing fields adopted are not geodesic on the background;
\item the Killing fields adopted do not have constant norm on the background; and
\item the three-dimensional subgroup adopted to define homogeneity is
  not a normal subgroup of the full isometry group, {\it i.e.} it does not
  map onto itself under the action of rotations.
\end{itemize}
To be clear, this lack of symmetry is in addition to the local
anisotropies which will be introduced into the metric; it refers to
the way such metric perturbations will be transported from a point
through the space. In spacetimes formed from such models, observers
armed with gyroscopes are able to detect directly the changing
alignment of the Bianchi component of the CMB (or other anisotropic
observables) as they move from location to location.  Such effects can
also be found in the VII$_0$ flat model, and in some closed Type-IX
models, but they are entirely unavoidable in the open Bianchi models.

\subsection{Definition of the basis tetrad}

To formulate the dynamical description, we introduce the invariant
triad $\vec{e}_i$ obeying
\begin{equation}
[\vec{e}_i,{\bm \xi}_j]=0 \, .\label{eq:invariant-basis}
\end{equation}
The $\vec{e}_i$ are not in general Killing but form a basis in which
any homogeneous tensor, in particular the metric, will have constant
components. The non-degeneracy of the basis at every point is
guaranteed by the transitivity of the action.

The $\vec{e}_i$ are computed by selecting any basis for the tangent
space at a point $p$ and dragging out with
equation~(\ref{eq:invariant-basis}). We can choose to orthogonalize
the $\vec{e}_i$ at $p$ (and hence everywhere) relative to the
background metric. We can further satisfy $\vec{e}_i|_p = \vec{{\bm
    \xi}}_i|_p$ by a suitable reparametrization of the ${\bm
  \xi}_i$. One then has $[\vec{e}_i,\vec{e}_j]=+C^k_{ij} \vec{e}_k$.
By making further linear reparametrizations of the $\vec{e}_i$ and
${\bm \xi}_i$, the $C^k_{ij}$ are brought into canonical form without
disturbing the orthogonality (see Table~\ref{tab:frw-bianchi}). We
have intentionally left the $\vec{e}_i$ un-normalized, so that, in the
maximally-symmetric background, $g(\vec{e}_i,\vec{e}_j) = e^{2 \alpha}
\delta_{ij}$ for some $\alpha$. Alternatively, one could set
$\alpha=0$ to gain an orthonormal triad, but the structure constants
cannot generally be chosen to be canonically normalized in this case.

Four-dimensional spacetimes with maximally-symmetric 3-spaces can be
constructed in the standard manner, by introducing a foliation of the
3-spaces labelled by a time $t$, and using the hypersurface-orthogonal
one-form $\tilde{\vec{n}}=\dd t$ to complete the spacetime metric:

\begin{equation}
g = - \tilde{\vec{n}} \otimes \tilde{\vec{n}} +
e^{2 \alpha(t)} \vec{e}^i \otimes\vec{e}^j \delta_{ij} \, ,
\label{eq:backgroundmetric}
\end{equation}
where $\{\vec{e}^i\}$ form the basis dual to $\{\vec{e}_i\}$. Both
$\{\vec{e}^i\}$ and $\{\vec{e}_i\}$ are time-invariant, {\it i.e.}\
their Lie derivative along $\vec{n}$ -- the vector corresponding to
$\tilde{\vec{n}}$ -- vanishes. The
tetrad $\{\vec{n},\vec{e}_i\}$ is known as the time-invariant tetrad
for the Bianchi spacetime~\cite{1969MNRAS.142..129H}.

\section{Killing fields and invariant vectors for specific spaces}\label{sec:coord-expr-modes}
\label{sec:coord_expressions}

In this section, we write explicit expressions for all six Killing
vector fields of each maximally-symmetric manifold, then perform the
decomposition into isotropy and homogeneity groups as described in
Section \ref{sec:what-homogeneity}.

This both elucidates the origin of the Bianchi symmetries and allows
construction of coordinate expressions for the Killing fields and
invariant basis.  While such expressions have been calculated by Taub
\cite{taub1951est} (see also Ryan and Shepley
\cite{1975hrc..book.....R}), they are presented in a form which
results in a zero-order metric not immediately familiar to
cosmologists. We will present our results here in as generally
applicable a form as possible; longer expressions which result from
writing the vector fields in familiar coordinate charts are given in
\ref{sec:altern-coord-expr}.

\subsection{Flat space: Bianchi Types I and VII$_0$}

\begin{figure}
\begin{flushright}
\includegraphics[width=0.9\textwidth]{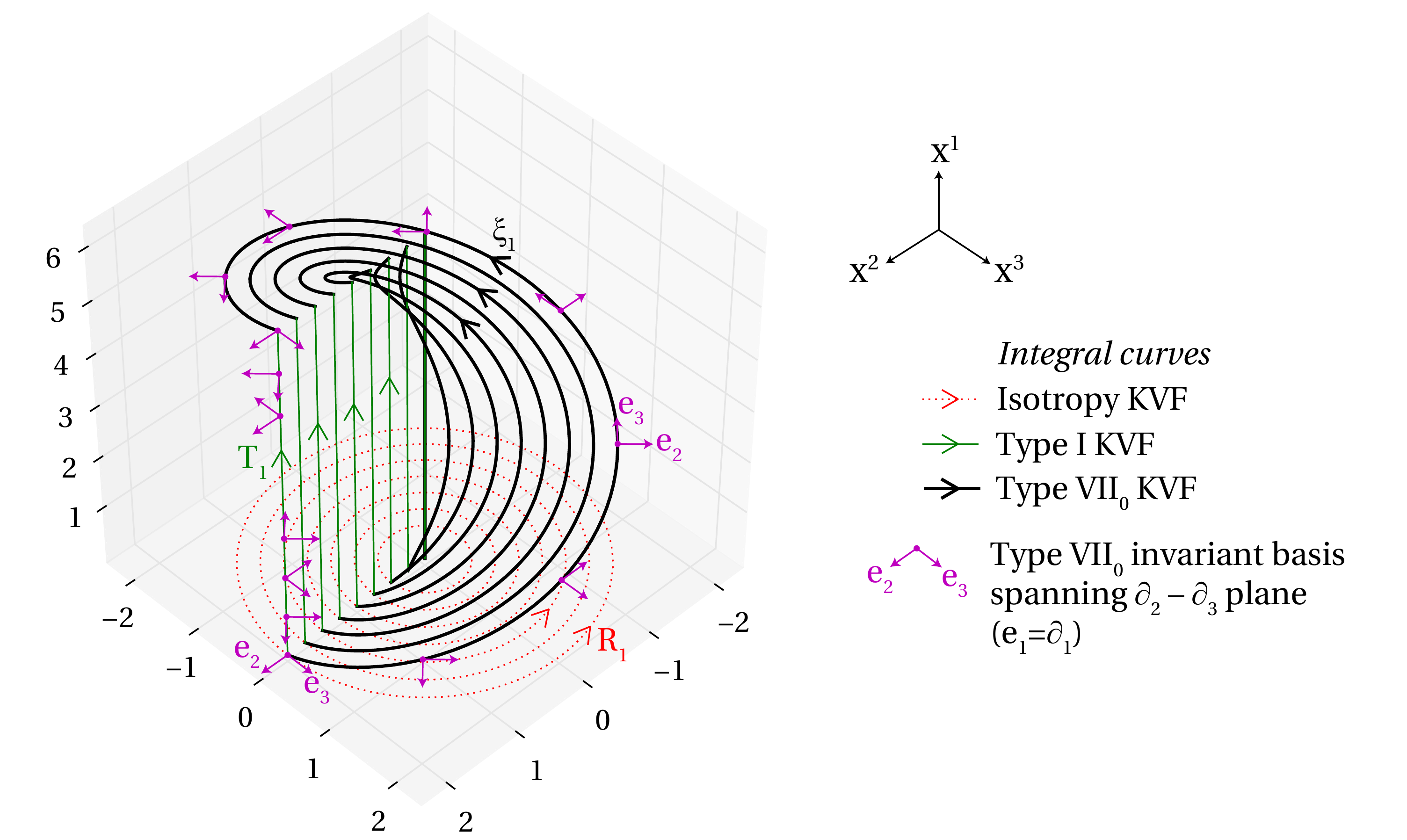}
\end{flushright}
\caption{The Killing fields of Type VII$_0$ compared to those of Type
  I; some integral curves of both are shown (as thick and thin solid
  lines respectively). All the Type-I Killing fields ${\bm \xi}^{I}_i$
  (along with ${\bm \xi}^{\mathrm{VII}_0}_2$ and ${\bm \xi}^{\mathrm{VII}_0}_3$)
  induce translations along the coordinate axes, whereas
  ${\bm \xi}^{\mathrm{VII}_0}_1$  (as a linear combination of $\partial_1$
  and $R_1$) induces a more complex spiralling motion. The vectors of
  the invariant triad $\vec{e}_2^{\mathrm{VII}_0}$, $\vec{e}_3^{\mathrm{VII}_0}$
  are also plotted. These spiral in response to the motion of the
  Killing fields; see equation
  (\ref{eq:basis-vii0}).}\label{fig:type_vii_kvfs}
\end{figure}

The six Killing vector fields of flat space described by Cartesian
coordinates $\{x^i\}$ are most simply expressed as the translations
$\vec{T}_i=\partial_i$ and the rotations $\vec{R}_i = \epsilon_{ijk}
x^j \partial_k$, with commutation relations

\begin{eqnarray}
\left[\vec{T}_i,\vec{T}_j \right]=0 \nonumber \\
\left[\vec{R}_i,\vec{R}_j \right] = -\epsilon_{ijk} \vec{R}_k \nonumber \\
\left[\vec{R}_i,\vec{T}_j \right] = -\epsilon_{ijk} \vec{T}_k \textrm{.}
\end{eqnarray}
Solution of the closure requirement~(\ref{eq:kvf-closure}) in terms of
the $\rho_i^j$~(\ref{eq:rot-coeffs}) leaves substantial freedom, but
up to reparametrizations the two possibilities obtained give the Type-I
and VII$_0$ KVFs (${\bm \xi}^{I}_i$ and ${\bm \xi}^{\mathrm{VII_0}}_i$ respectively) in
canonical form:
\begin{eqnarray}
{\bm \xi}^{I}_i = \vec{T}_i 
\end{eqnarray}
and
\begin{eqnarray}
{\bm \xi}^{\mathrm{VII_0}}_1 = \vec{T}_1 + \vec{R}_1 \label{eq:I-becomes-VII0} \nonumber \\
{\bm \xi}^{\mathrm{VII_0}}_j = \vec{T}_j \hspace{2cm} j \ne 1 \textrm{.}
\end{eqnarray}
The $\vec{T}_i$ have constant norm, form a subalgebra which is also
closed under rotations and have geodesic integral curves; as noted in
Section~\ref{sec:what-homogeneity}, these conditions are all
equivalent. The Type-VII$_0$ KVFs do not satisfy such conditions,
because integral curves of the ${\bm \xi}_1$ fields describe a
spiralling motion (see Figure~\ref{fig:type_vii_kvfs}).

Because the background is flat, parallel transport of homogeneous
quantities between different points on the manifold is uniquely
defined; it may be verified that in the Type-I (but not VII$_0$) case
the Lie transport coincides with parallel transport.

In the Type-I case, $\vec{T}_i$ are Abelian and so form their own
reciprocal group. For Type VII$_0$, the reciprocal group is given by
solving $[\vec{e}_i,{\bm \xi}_j]=0$ subject to the constraint $\vec{e}_i|_{x=0} =
\vec{{\bm \xi}}_i|_{x=0}$:

\begin{equation}
\left(\begin{array}{c} \vec{e}^{\mathrm{VII_0}}_1 \\
\vec{e}^{\mathrm{VII_0}}_2 \\
\vec{e}^{\mathrm{VII_0}}_3 \end{array}\right) = 
\matr{1 & 0 & 0 \\
0 & \cos x^1 & \sin x^1 \\
0 & -\sin x^1 & \cos x^1 }
\left(\begin{array}{c} \partial_1 \\
\partial_2 \\
\partial_3 \end{array}\right) \textrm{.}\label{eq:basis-vii0}
\end{equation}
These solutions are illustrated by plotting the vectors
$\vec{e}^{\mathrm{VII_0}}_2$ and $\vec{e}^{\mathrm{VII_0}}_3$ at
various points in Figure \ref{fig:type_vii_kvfs}. Recalling that
anisotropic quantities in the Type-VII$_0$ case will have constant
components relative to the $\vec{e}^{\mathrm{VII_0}}_i$ basis, the
origin of spiralling patterns in such universes becomes clear from
inspecting this diagram.

\subsection{Open space: Bianchi Types V and VII$_h$}\label{sec:open-space}

\begin{figure}
\begin{flushright}
\includegraphics[width=1.0\textwidth]{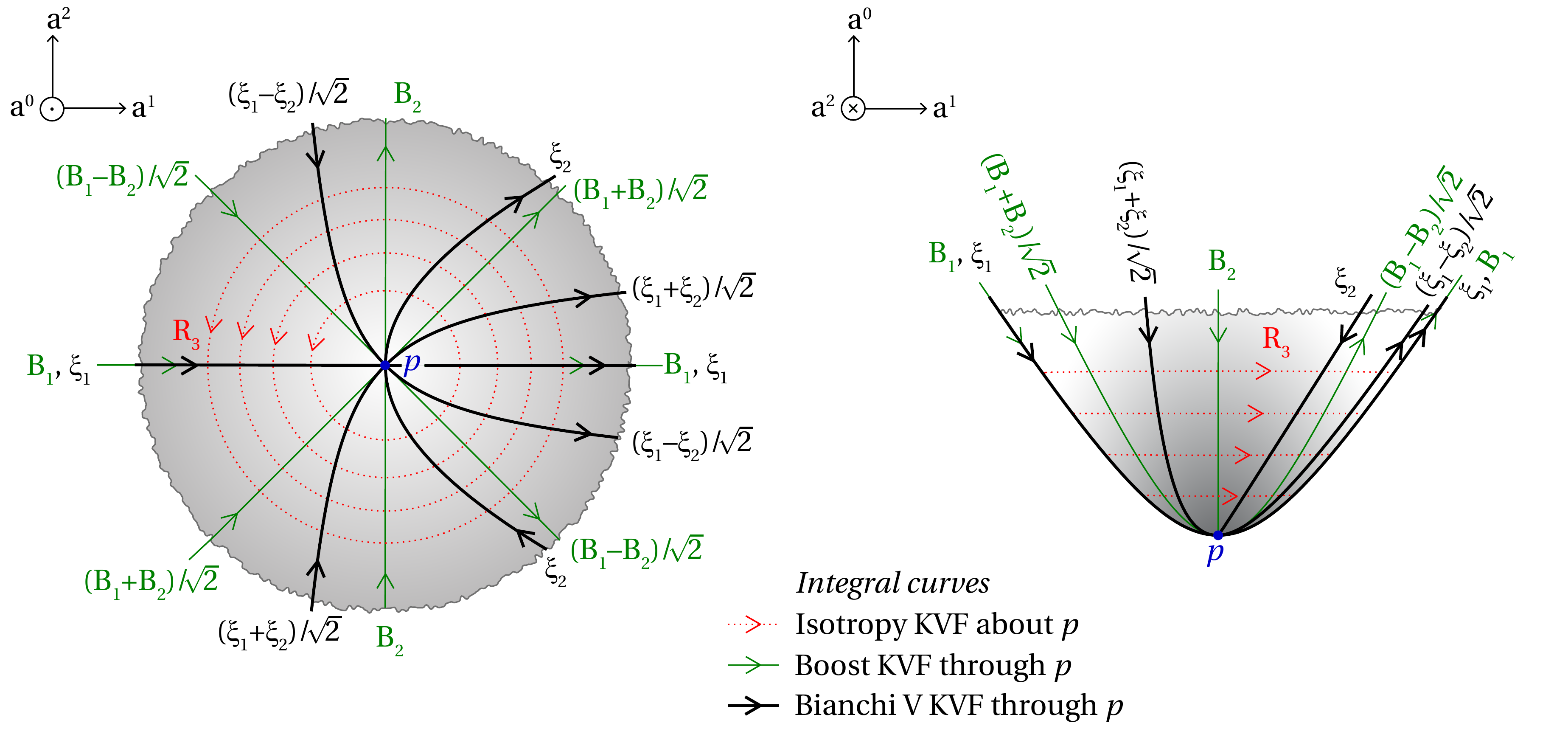}
\end{flushright}
\caption{The open space as a hyperboloid embedded in Minkowski space;
  here a cut is taken along $a^3=0$, and a portion of the resulting
  geometry illustrated in two projections.  In particular we plot
  integral curves of KVFs, decomposed about the point $p$ (where
  $a^0=1$) as described in the text.  The rotational motion is shown
  as a dotted line, while boosts are shown as thin solid
  lines. However, the boosts do not form a closed group and thus do
  not qualify as generators of homogeneity according to the criteria
  in Section \ref{sec:what-homogeneity}. Thus the Bianchi models take
  a linear combination of the boosts and rotations to form the basis
  set of Bianchi KVFs at $p$ (see equation~\ref{eq:KVF-v}). Here we have
  plotted integral curves of the Bianchi Type-V motions (heavy solid lines)
  which agree with the boosts in the tangent space of $p$, but deviate
  on scales of order the curvature radius. The rotational asymmetry of
  these integral curves, which is unavoidable in open models,
  ultimately leads to the non-trivial appearance of the Bianchi
  CMB. }\label{fig:type_v_kvfs}
\end{figure}

To obtain coordinate expressions for the Killing and basis vectors, we
consider the well-known embedding of an open manifold in Minkowski
space.  The ambient metric is

\begin{equation}
\dd s^2 =  - \dd (a^0)^2 + \dd (a^1)^2 + \dd (a^2)^2 + \dd (a^3)^2 \textrm{.}\label{eq:open-ambient}
\end{equation}
The hypersurface representing the open space, with curvature set to
$1$ without loss of generality, is defined by
\begin{equation}
1 =  (a^0)^2 - \left[(a^1)^2+(a^2)^2+(a^3)^2\right] \textrm{.} \label{eq:open-hypersurface}
\end{equation}
The group of continuous global linear transformations which preserve
(\ref{eq:open-ambient},~\ref{eq:open-hypersurface}) is the restricted
Lorentz group SO$^{+}(3,1)$. There are six Killing vector fields which
can be decomposed about a point $p$ into three rotations ($\vec{R}_i$)
and three boosts ($\vec{B}_i$):
\begin{eqnarray}
\vec{R}_i = \epsilon_{ijk} a^j \partial_k \nonumber \\
\vec{B}_i = a^i \partial_0 + a^0 \partial_i\textrm{,}
\end{eqnarray}
where the special point is taken without loss of generality to be
$a^0=1$, $a^i=0$ ($i>0$). These Killing fields have well-known
commutators:
\begin{eqnarray}
  \left[ \vec{R}_i,\vec{R}_j \right]  & = & - \epsilon_{ijk}\vec{R}_k \nonumber \\
  \left[ \vec{B}_i,\vec{B}_j \right]  & = & \epsilon_{ijk}\vec{R}_k  \nonumber \\
  \left[ \vec{R}_i,\vec{B}_j \right]  & = & - \epsilon_{ijk}\vec{B}_k \label{eq:open-commu}\textrm{.}  
\end{eqnarray} 
The commutation relations (\ref{eq:open-commu}) show that the boosts
$\vec{B}_i$ are not suitable generators for homogeneity because they
do not form a closed group. One must therefore consider linear
combinations of the form (\ref{eq:rot-coeffs}). It may be verified
that the only closed possibilities (unique up to transformations under
the isotropy subgroup $\{\vec{R}_i\}$) give either a realisation of
the canonical Type-V or VII$_h$ KVFs.\footnote{For both Types V and
  VII$_h$, the normalization of $\vec{{\bm \xi}}_j$ for $j \ne 1$ can be
  changed independently of the normalization of $\vec{{\bm \xi}}_1$ while
  still leaving the structure constants in canonical form. In the
  VII$_h$ case the scaling of ${\bm \xi}_2$ must be equal to that of
  ${\bm \xi}_3$. Here we have chosen normalizations which fit in with our
  formalism, in that they lead to a spatial metric $\propto
  \delta_{ij}$ in the exactly isotropic limit, {\it cf.} equation
  (\ref{eq:backgroundmetric}).}  In the former case we have
\begin{equation}
  \vec{{\bm \xi}}^{\mathrm{V}}_1 = \vec{B}_1 , \qquad
\vec{{\bm \xi}}^{\mathrm{V}}_2 = \vec{B}_2 - \vec{R}_3 , \qquad
\vec{{\bm \xi}}^{\mathrm{V}}_3 = \vec{B}_3 + \vec{R}_2 \label{eq:KVF-v}\textrm{.}
\end{equation}
The effect of this choice is shown in Figure \ref{fig:type_v_kvfs} by
displaying the hyperboloid described in the space spanned by the $a^i$
coordinates (taking $a^3=0$). In particular, we have illustrated
integral curves through $p$ of both the $\vec{B}_1$, $\vec{B}_2$ subset (thin
solid lines) and the ${\bm \xi}_1$, ${\bm \xi}_2$ subset (heavy solid lines). Both
of these sets span the $a^1, a^2$ tangent space at $p$ which allows
for a direct comparison. While the $\vec{B}_i$ subset is closed under
rotations, this is
not the case for the ${\bm \xi}_i$ subset; for instance, it is clear from
the figure that ${\bm \xi}_2$ is not obtained from any rotation about $p$ of
${\bm \xi}_1$. We emphasize again that, despite their attractive qualities,
the $\vec{B}_i$ are not suitable for transporting anisotropic quantities
since they do not form a closed subalgebra (Section
\ref{sec:what-homogeneity}).

For Type VII$_h$ one has
\begin{eqnarray}
  \vec{{\bm \xi}}^{\mathrm{VII}_h}_1 =\sqrt{h}\, \vec{{\bm \xi}}_1^{\mathrm{V}} +  \vec{R}_1  \nonumber \\
  \vec{{\bm \xi}}^{\mathrm{VII}_h}_j = \sqrt{h}\, \vec{{\bm \xi}}^{\mathrm{V}}_j \hspace{3cm} (j \ne 1) \label{eq:v-becomes-viih}\textrm{.}
\end{eqnarray}
Note the similarity of the relation (\ref{eq:I-becomes-VII0})
between Type-I and VII$_0$ Killing fields with the relation
(\ref{eq:v-becomes-viih}) between Type V and VII$_h$. Both introduce a
`spiralling' motion into one of the Killing fields. In the flat case,
this is fixed (in the canonical decomposition) to have a unit comoving
scale length which is later adjusted in physical scale by the
choice of spatial scale-factor at redshift zero. In the open case, there is an
additional freedom in the relative scale of the isotropic curvature
and spiral motion; this is fixed by adjusting the value of the structure constant
$\sqrt{h}$.

We can find invariant vectors by solving $[\vec{{\bm \xi}}_i,\vec{e}_j]=0$
subject to the constraint $\vec{e}_i = {\bm \xi}_i$ at
$a^0=1$. For Type V, we find
\begin{eqnarray}
\vec{e}^{\mathrm{V}}_1 &=& \left(e^x -a^0\right) \partial_0
+ \left(e^x -a^1\right) \partial_1 - a^2 \partial_2
- a^3 \partial_3 \nonumber \\
\vec{e}^{\mathrm{V}}_2 &=& e^x {\bm \xi}^{\mathrm{V}}_2 \nonumber \\
\vec{e}^{\mathrm{V}}_3 &=& e^x {\bm \xi}^{\mathrm{V}}_3 ,
\label{eq:inv_V}
\end{eqnarray}
where $x = - \ln (a^0 -a^1)$. (For a geometrical interpretation of
$x$, see \ref{subsec:open_app}.)  For VII$_h$, a suitable invariant
basis is
\begin{equation}
\left(\begin{array}{c} \vec{e}^{\mathrm{VII}_h}_1 \\
\vec{e}^{\mathrm{VII}_h}_2 \\
\vec{e}^{\mathrm{VII}_h}_3 \end{array}\right) = 
\sqrt{h} \matr{1 & 0 & 0 \\
0 & \cos (x/\sqrt{h}) & \sin (x/\sqrt{h}) \\
0 & -\sin (x/\sqrt{h}) & \cos (x/\sqrt{h}) }
\left(\begin{array}{c} \vec{e}^{\mathrm{V}}_1 \\
\vec{e}^{\mathrm{V}}_2 \\
\vec{e}^{\mathrm{V}}_3 \end{array}\right) . \label{eq:link-V-to-VIIh-invariant}
\end{equation}
This simple formulation
arises because the ${\bm \xi}_i$ are closed under the action of $\vec{R}_1$,
so the Jacobi identities applied to
$0=[\vec{R}_1,[{\bm \xi}_j,\vec{e}_k]]$
may be used to show that the
$\{\vec{e}^{V}_i\}$ are closed under the action of $\vec{R}_1$. The
corresponding commutation coefficients are fixed by the behaviour of
$\vec{R}_i$ on the tangent space at the identity, {\it i.e.}
$[\vec{R}_1,\vec{e}_1^{\mathrm{V}}]=0$,
$[\vec{R}_1,\vec{e}_2^{\mathrm{V}}]=-\vec{e}^{\mathrm{V}}_3$,
$[\vec{R}_1,\vec{e}_3^{\mathrm{V}}]=\vec{e}^{\mathrm{V}}_2$. (These
may be verified directly from equation~\ref{eq:inv_V}.)
The
construction is therefore completely analogous to that of the VII$_0$
invariant fields.

\subsection{Closed space: Bianchi type IX}\label{sec:clos-spac}

The closed case has been discussed extensively in Ref. 
\cite{1991PhRvD..44.2356K}.  A suitable embedding may be
considered with the ambient metric and surface function
\begin{eqnarray}
\dd s^2 = (\dd a^0)^2 + (\dd a^1)^2 + (\dd a^2)^2 + (\dd a^3)^2 \nonumber \\
1 = (a^0)^2 + (a^1)^2+ (a^2)^2 + (a^3)^2 \textrm{,}
\end{eqnarray}
with KVFs
\begin{eqnarray}
\vec{R}_i = \epsilon_{ijk}a^j \partial_k \nonumber \\
\vec{T}_i = a^i \partial_0 - a^0 \partial_i \textrm{.}
\end{eqnarray}
The commutation relations are
\begin{eqnarray}
\left[\vec{R}_i,\vec{R}_j\right] = - \epsilon_{ijk} \vec{R}_k \nonumber \\
\left[\vec{T}_i,\vec{T}_j\right] = - \epsilon_{ijk} \vec{R}_k\label{eq:sign-reversed} \nonumber \\
\left[\vec{R}_i,\vec{T}_j\right] = -\epsilon_{ijk} \vec{T}_k\textrm{.}
\end{eqnarray}
In the vicinity of $a^0=1$, $a^i=0$, the $\vec{T}_i$ correspond to
Euclidean translations along the $i$th coordinate direction and
the $\vec{R}_i$ correspond to rotations about $a^0=1$, $a^i=0$. 

There are two simply-transitive subgroups
\begin{equation}
{\bm \xi}^{\pm}_i = \frac{1}{2}\left(\vec{T}_i \pm \vec{R}_i\right), \label{eq:kvf-ix}
\end{equation}
and both yield Type-IX structure constants:
\begin{equation}
\left[{\bm \xi}^{\pm}_i,{\bm \xi}^{\pm}_j\right] = \mp \epsilon_{ijk} {\bm \xi}^{\pm}_k .
\end{equation}
With the factor $1/2$ in (\ref{eq:kvf-ix}), the ${\bm \xi}^{+}_i$ have
canonical commutation relations, while reversing the sign of the ${\bm
  \xi}^{-}_i$ makes this set canonical too.  One may verify that both
sets of Killing vector fields have constant norm on the surface, so
are tangent to congruences of geodesics as described for the flat case
above. From this it follows that parallel transport and Lie transport
along a Killing field can differ only by a rotation in the plane
perpendicular to the transport direction. This behaviour can
succinctly account for the appearance of the CMB in Type-IX models:
the temperature pattern is a pure quadrupole but a screen-projected
polarization vector spirals at a constant rate in conformal time
\cite{2009PhRvD..79j3518P}. (However, for an exception to this
behaviour see the near-VII$_0$ models in
Section~\ref{sec:other-bianchi-types}.)

Since $[{\bm \xi}^{+}_i,{\bm \xi}^{-}_j]=0$, the invariant fields are also Killing
(the metric is bi-invariant; it is, actually, the
Killing metric for this space). For consistency we should adopt
${\bm \xi}^{\mathrm{IX}}_i = {\bm \xi}^{+}_i$,
$\vec{e}^{\mathrm{IX}}_i={\bm \xi}^{-}_i$, although these roles are reversed
under parity.

\section{Enumerating the linearised Bianchi modes}\label{sec:enum-line-bianchi}

In this section, we present a derivation of the independent modes
which could arise in an almost-isotropic Universe consistent with
Bianchi-type symmetries.

Our starting point is the metric in equation~(\ref{eq:backgroundmetric}).
We can break isotropy while preserving homogeneity by adding a spatial metric
perturbation of the form
\begin{equation}
\delta g = \delta{g}_{ab}(t)\vec{e}^a \otimes\vec{e}^b \, .
\end{equation}
As described in Section \ref{sec:what-homogeneity}, the perturbed
metric is invariant under the action of the homogeneity group, which
is a subgroup of the original isometries for the unperturbed FRW
spacetime.  The full, perturbed metric can be written in the form \cite{1969MNRAS.142..129H,1968ApJ...151..431M}
\begin{equation}
g =  -\tilde{\vec{n}} \otimes \tilde{\vec{n}} + e^{2 \alpha(t)} (e^{2 \beta(t)})_{ab} \vec{e}^a \otimes \vec{e}^b \, ,
\label{eq:perturbmetric}
\end{equation}
where the symmetric, trace-free (STF) $\beta_{ab}(t)$ and, generally, $\alpha(t)$ are perturbed from
their FRW values. By construction, the isotropic limit corresponds to
$\beta_{ab} = \dot{\beta}_{ab} = 0$, where an overdot denotes the
derivative with respect to $t$.  The evolution of $\beta$ contains all
information about the anisotropic perturbation (at any order).

The evolution (Einstein) equations are most simply expressed relative
to an orthonormal frame, denoted $\hat{\vec{e}}_{\hat{a}}$. One may
construct such a frame explicitly \cite{1969MNRAS.142..129H} via the transformation
\begin{equation}
\hat{\vec{e}}_{\hat{a}} = e^{-\alpha} \left( e^{-\beta}\right)_{\hat{a}a} \vec{e}_a \textrm{.}
\end{equation}
The expansion $H$, shear $\sigma_{\hat{a}\hat{b}}$ and
Fermi-relative rotation $\Omega_{\hat{a}}$ of the orthonormal
frame\footnote{Note that $\nu_{\hat{a}\hat{b}} =
  \epsilon_{\hat{a}\hat{b}\hat{c}}\Omega^{\hat{c}}$ for comparison
  with certain works such as those by Hawking and
  Collins~\protect\cite{1973ApJ...180..317C,1969MNRAS.142..129H,1973MNRAS.162..307C}.}
are defined by
\begin{equation}
\vec{e}_{\hat{a}} \cdot [\vec{n}, \vec{e}_{\hat{b}}] = -H \delta_{\hat{a}\hat{b}} -\sigma_{\hat{a}\hat{b}} -\epsilon_{\hat{a}\hat{b}\hat{c}} \Omega^{\hat{c}} \textrm{,}
\end{equation}
where $\sigma_{\hat{a}\hat{b}}$ is symmetric and trace-free. We shall
now linearise about the background FRW solution by requiring
\begin{eqnarray}
  \left| \beta_{ab} \right| & =  \mathcal{O}(\epsilon)  \nonumber \\
  \left| \dot{\beta}_{ab}/H \right| & =  \mathcal{O}(\epsilon) \nonumber \\
e^{-2\alpha} \left| \beta_{ab}  C^2/H^2 \right| & =  \mathcal{O}(\epsilon) \, ,
\label{eq:linear}
\end{eqnarray}
where $\epsilon\ll 1$. The final condition can be interpreted as
putting a lower bound on the wavelength of the perturbations as a
fraction of the horizon size, which is necessary for spatial derivatives
of perturbed quantities to remain small. We
perform a fiducial fluid decomposition of the stress tensor as
\begin{equation}
T_{\mu\nu} = \rho n_{\mu} n_{\nu} + p (g_{\mu\nu} + n_{\mu} n_{\nu}) + 2 n_{(\mu} P_{\nu)} + \pi_{\mu\nu} ,
\end{equation}
where $\rho$
is the fluid energy density measured by observers comoving with the $\vec{n}$
congruence, $p$ the isotropic pressure, $P_\mu$ the momentum density
and $\pi_{\mu\nu}$ the anisotropic stress. Here $P_\mu$ and $\pi_{\mu\nu}$ are
orthogonal to $\vec{n}$ ($P_{\mu} n^{\mu}=\pi_{\mu\nu} n^{\nu}=0$)
and $\pi_{\mu\nu}$ is symmetric and trace-free. We further
assume
\begin{equation}
\left| \pi_{\mu\nu}/H^2 \right| = \mathcal{O}(\epsilon) , \qquad
 \left| P_{\mu}/H^2 \right| = \mathcal{O}(\epsilon) \, ,
\end{equation}
and will in practice shortly restrict $\pi_{\mu\nu}=0$, {\it i.e.} a perfect
fluid, although unlike many Bianchi model analyses we will allow for
small non-zero momentum density $P_{\mu}$ (corresponding to tilt of
the fluid source).

The first-order kinematic and dynamical quantities in the orthonormal
frame are
\begin{eqnarray}
e^{\alpha} \hat{a}_{\hat{a}} & = a_{\hat{a}} - \beta_{\hat{a}a} a_a + \mathcal{O}(\epsilon^2)  \\
e^{\alpha} \hat{n}^{\hat{a}\hat{b}} & = n^{\hat{a}\hat{b}} + \beta_{\hat{a}a} n^{a\hat{b}} + \beta_{\hat{b}b} n^{\hat{a}b} + \mathcal{O}(\epsilon^2)  \\
\sigma_{\hat{a}\hat{b}}/H & = \dot{\beta}_{\hat{a}\hat{b}}/H + \mathcal{O}(\epsilon^2) \\
H & = \dot{\alpha}  \\
\Omega_{\hat{c}}/H & = \mathcal{O}(\epsilon^2) \textrm{.}
\label{eq:metric-shear}
\end{eqnarray}
These equations are non-tensorial (valid only in the orthonormal frame); the components of $a$ and $n$ on the right-hand sides are the
canonical, time-invariant structure constants and the components of
$\beta_{ab}$ are as defined in the time-invariant frame.

The expansions above may be inserted directly into the orthonormal-frame Einstein equations (see, e.g., Ref. \cite{wainwright1997dsc}; or
for a somewhat different form, Ref. \cite{1973MNRAS.162..307C}),
giving
\begin{eqnarray}
\rho  & = 3 H^2 + \tensor[^{3}]{R}{}/2 -\Lambda   + \mathcal{O}(H^2 \epsilon^2) & \label{eq:friedmann} \\
\dot{H} & = - H^2 - (\rho + 3 p)/6 +\Lambda/3  + \mathcal{O}(H^2 \epsilon^2) & \label{eq:raychaudhuri}\\
\ddot{\beta}_{\hat{a}\hat{b}} &  = -3H \dot{\beta}_{\hat{a}\hat{b}} - \tensor[^3]{S}{_{\hat{a}\hat{b}}} + \pi_{\hat{a}\hat{b}} + \mathcal{O}(H^2 \epsilon^2) & \label{eq:shear-evolution} \\
P_{\hat{a}} & =  e^{-\alpha} \left(3\dot{\beta}_{\hat{a}b} a_{b} - \epsilon_{\hat{a}bc} \dot{\beta}_{bd} n_{dc} \right)
 +\mathcal{O}(e^{-\alpha} HC\epsilon^2) \textrm{,} \label{eq:tilt-constraint}
\end{eqnarray}
with the fluid conservation equation
\begin{equation}
\dot{\rho} = -3H(\rho+p) + 2 e^{-\alpha} a_{\hat{a}} P^{\hat{a}} + \mathcal{O}(e^{-2\alpha}HC^2\epsilon^2, H^3 \epsilon^2)\textrm{,}
\end{equation}
where the structure constants $a$ and $n$ again take their canonical
numerical values.  In equations (\ref{eq:friedmann}) and
(\ref{eq:raychaudhuri}), $\tensor[^3]{R}{}$ and
$^{3}S_{\hat{a}\hat{b}}= \tensor[^{3}]{R}{_{\hat{a}\hat{b}}} -
\tensor[^3]{R}{}\delta_{\hat{a}\hat{b}}/3 $ are the isotropic and
anisotropic parts of the spatial 3-curvature respectively:
\begin{eqnarray}
\hspace{-1cm} ^{3}R e^{2\alpha} & = & \tensor[^{3}]{R}{^{\hat{a}}_{\hat{a}}} e^{2 \alpha} =   -n_{ab} n_{ab} + \frac{1}{2} n_{aa} n_{bb} -6 a_{a} a_{a} \nonumber \\
& & \hspace{2cm} + \beta_{ac} (-4 n_{ab} n_{cb} + 2 n_{ac} n_{bb} +12 a_a a_c)  + \mathcal{O}(C^2 \epsilon^2) \label{eq:iso-curv-expand} 
\end{eqnarray}
\begin{eqnarray}
\hspace{-1cm}\tensor[^{3}]{S}{_{\hat{a}\hat{b}}}e^{2\alpha}  & = &  2 n_{\ispace c\langle\hat{a}} n_{\hat{b}\rangle c} - n_{cc} n_{\langle\hat{a}\hat{b}\rangle} - 2 \epsilon_{\ispace cd(\hat{a}} n_{\hat{b})c} a\indices{_d} \nonumber  \\*
& &  +  2\left[-\epsilon_{\ispace cd(\hat{a}} \beta_{\hat{b})e} n_{ec} a_d - \epsilon_{\ispace cd(\hat{a}} n_{\hat{b})e} a_d \beta_{ec} 
+\epsilon_{\ispace cd(\hat{a}}n_{\hat{b})c}a_e \beta_{ed} \right. + 2 \beta_{cd} n_{\ispace c\langle\hat{a}}n_{\hat{b}\rangle d} \nonumber \\
& & \left. \hspace{0.9cm} + 2\beta_{\ispace d\langle\hat{a}} n_{\hat{b}\rangle c} n_{cd} -
\beta_{cd}n_{cd}n_{\langle\hat{a}\hat{b}\rangle} - n_{cc} n_{\ispace d\langle\hat{a}}\beta_{\hat{b}\rangle d} \right]
+\mathcal{O}(C^2 \epsilon^2)  \textrm{,} \label{eq:aniso-curv-expand}
\end{eqnarray}
where angle brackets ($\langle \rangle$) denote symmetrising and removing the trace
over the enclosed indices.

\begin{table}[t!]
\begin{center}
\begin{tabular}{r|c|cc|cc}
\newcommand\tspat{\rule{0pt}{2.6ex}}
\newcommand\B{\rule[-1.2ex]{0pt}{0pt}}
       & $s$ & $v_1$ & $v_2$ & $t_1$ & $t_2$ \\[0.5ex]
  \hline & & & & &  \\[-2.0ex]
    {\bf VII$_0$} \rule{0pt}{3.ex}  & \matrthree{2}{0}{0}{-1}{0}{-1} & \matrthree{0}{1}{0}{0}{0}{0} & \matrthree{0}{0}{1}{0}{0}{0}  &  \matrthree{0}{0}{0}{1}{0}{-1} & \matrthree{0}{0}{0}{0}{1}{0}  \\
  $\mathcal{S}=$ & 0 & 0 & 0 & 4 & 4 \\
  $\mathcal{R}=$ & 0 & 0 & 0 & 0 & 0 \\
  $\vec{P}=$ & Transverse & \vecthree{0}{0}{-1} & \vecthree{0}{1}{0} & \multicolumn{2}{c}{Transverse} \\[0.4cm]
  \hline & & & & &  \\[-2.0ex]
  {\bf V} \rule{0pt}{3.ex}& \matrthree{2}{0}{0}{-1}{0}{-1} & \matrthree{0}{1}{0}{0}{0}{0} & \matrthree{0}{0}{1}{0}{0}{0}  &  \matrthree{0}{0}{0}{1}{0}{-1} & \matrthree{0}{0}{0}{0}{1}{0}  \\
  $\mathcal{S}=$ & 0 & 0 & 0 & 0 & 0 \\
  $\mathcal{R}=$&$24$ & 0 & $0$ & 0 & 0 \\
  $\vec{P}=$ & \vecthree{6}{0}{0} & \vecthree{0}{3}{0} & \vecthree{0}{0}{3}  & \multicolumn{2}{c}{Transverse} \\[0.4cm]
  \hline  & & & & &  \\[-2.0ex]
  {\bf VII$_h$} \rule{0pt}{3.ex} & \matrthree{2}{0}{0}{-1}{0}{-1} & \matrthree{0}{1}{0}{0}{0}{0} & \matrthree{0}{0}{1}{0}{0}{0}  & \matrthree{0}{0}{0}{1}{i}{-1} & \matrthree{0}{0}{0}{1}{-i}{-1} \\
  $\mathcal{S} =$ & $0$ & 0 & 0 & $4(1-i\sqrt{h})$ & $4(1+i\sqrt{h})$ \\
  $\mathcal{R}=$&$24h$ & 0 & 0 & 0 & 0 \\
  $\vec{P}=$& \vecthree{6 \sqrt{h}}{0}{0} & \vecthree{0}{3\sqrt{h}}{-1} & \vecthree{0}{1}{3 \sqrt{h}} & \multicolumn{2}{c}{Transverse} \\[0.4cm]
  \hline \\[-2.0ex]
  {\bf I}& \multicolumn{5}{l}{Any STF $\beta$ has $\mathcal{S}=0$, $\mathcal{R}=0$. All modes transverse.} \\
  {\bf IX} & \multicolumn{5}{l}{Any STF $\beta$ has $\mathcal{S}=2$, $\mathcal{R}=0$. All modes transverse.}  \\[0.5ex]
  \hline
\end{tabular}
\end{center}
\caption{The decoupled linearized modes of Bianchi perturbations about
  an FRW Universe, expressed as the perturbation to the spatial metric
  relative to the $\{\vec{e}_a\}$ basis. Under each mode, the
  eigenvalue for the anisotropic curvature ($\mathcal{S}$) is given
  along with the perturbation to the isotropic curvature
  ($\mathcal{R}$) and momentum-density constraint ($\vec{P}$) as
  described by equations (\ref{eq:eigenmode}),
  (\ref{eq:R-perturbation}) and (\ref{eq:eigenmode-tilt})
  respectively. For transverse modes, the momentum density is
  zero. }\label{tab:frw-perturb}
\end{table}

By construction the zero-order contribution to the curvature arises
from the FRW background, with Ricci scalar values $R_{\mathrm{FRW}}$
listed in Table \ref{tab:frw-bianchi}. The anisotropic curvature to
first order is linear in $\beta$ so that one may write
\begin{equation}
\tensor[^{3}]{S}{_{\hat{a}\hat{b}}}e^{2\alpha} = \mathcal{S}_{cd\hat{a}\hat{b}} \beta_{cd}, \qquad \tensor[^{3}]{R}{}e^{2\alpha} = \tensor[^{3}]{R}{_{\mathrm{FRW}}}
e^{2\alpha} + \mathcal{R}_{\hat{a}\hat{b}} \beta_{\hat{a}\hat{b}} \textrm{,}
\end{equation}
and equation (\ref{eq:shear-evolution}) becomes
\begin{equation}
\ddot{\beta}_{\hat{a}\hat{b}} + 3 H \dot{\beta}_{\hat{a}\hat{b}} +  e^{-2\alpha} \mathcal{S}_{cd\hat{a}\hat{b}} \beta_{cd} - \pi_{\hat{a}\hat{b}} = 0 \textrm{.} \label{eq:evol-in-beta}
\end{equation}
Anisotropic stresses could arise, for example, from collisionless
free-streaming, magnetic fields or topological defects; see
Ref. \cite{1997PhRvD..55.7451B}. Their effect will be dependent on the
source evolution equations for $\pi_{\mu\nu}$; by combining such
expressions with our present formalism, many existing analyses of
anisotropic stresses could be extended from the simple Type-I setups
usually employed in such studies.  However, for the present work we
will assume a perfect fluid ($\pi_{\hat{a}\hat{b}}=0$); then the
Bianchi perturbations decouple into independent eigenmodes satisfying
\begin{equation}
\mathcal{S}_{ab\hat{c}\hat{d}}\beta^{(m)}_{ab} = \mathcal{S}^{(m)} \beta^{(m)}_{ab} \delta_{a\hat{c}} \delta_{b\hat{d}} \label{eq:eigenmode}
\end{equation}
where the $\beta^{(m)}$ are listed for each type in Table
\ref{tab:frw-perturb}. The split into three classes ($s$, $v$ and $t$) reflects
a rotational symmetry of the structure constants about the $\vec{e}_1$
axis.  The modes transform irreducibly as scalars ($s$; spin-0),
vectors ($v$; spin-1) and tensors ($t$; spin-2) under this
symmetry;\footnote{This decomposition can be made in
the non-linear case \protect\cite{2005CQGra..22..579C} and is an
example of the automorphism approach to Bianchi
dynamics (e.g. \protect\cite{1988PhR...166...89R} and references therein).}
these evolve independently in our linear approximation.

The appearance of complex eigenmodes for Type-VII$_h$ $t$ modes
indicates a time-dependent rotation of the mode expansion relative to
the $\vec{n}$-induced point identification between time slices. Real
physical quantities are obtained by combining the complex conjugate
mode pairs in linear combinations in the initial conditions; one may
verify that the reality is conserved by the time evolution.

A general Bianchi linear perturbation is written
\begin{equation}
\beta_{ab} = \sum_{m=1}^5 A^{(m)}(t) \beta_{ab}^{(m)} , \qquad
\sigma_{\hat{a}\hat{b}} = \sum_{m=1}^5 \dot{A}^{(m)}(t) \beta_{\hat{a}\hat{b}}^{(m)},
\label{eq:eigenmode-decomposition}
\end{equation}
with the amplitude of each mode evolving independently as
\begin{equation}
  A^{(m)\hspace{0.5mm}\prime\prime} + 2\mathcal{H} A^{(m)\hspace{0.5mm}\prime} + \mathcal{S}^{(m)} A^{(m)} = 0 \label{eq:eigenmode-evol} \textrm{.}
\end{equation}
Here we have made use of conformal time $\eta$; primes ($'$) denote
$\dd/\dd \eta=e^{\alpha} \dd / \dd t$ and $\mathcal{H}=H e^{\alpha} =
\alpha '$. Those modes with $\mathcal{S} \ne 0$ describe damped
harmonic oscillators and in general undergo oscillations, continually
exchanging anisotropy between shear and curvature as they die away.
The shear of other modes ($\mathcal{S}=0$) decays as $e^{-3 \alpha}$;
these modes also admit an $A^{(m)}=\mbox{constant}$ solution which is
a pure gauge artefact (see Section~\ref{sec:gaugemodes}).

The tilt constraint, equation~(\ref{eq:tilt-constraint}), reads
\begin{equation}
P_{\hat{a}}  =  e^{-\alpha} \sum_{m=1}^5 \dot{A}^{(m)} P^{(m)}_{\hat{a}} \textrm{,}\label{eq:eigenmode-tilt} \\
\end{equation}
where
\begin{equation}
 P^{(m)}_{\hat{a}}  =  3 \beta^{(m)}_{\hat{a}b}a_b - \epsilon_{\hat{a}bc} \beta^{(m)}_{bd} n_{dc} \, . 
\end{equation}
The components $P^{(m)}_{\hat{a}}$ are listed for each mode in 
Table \ref{tab:frw-perturb}. Those modes with vanishing momentum
density are labelled transverse in the table. The projected divergences
of the shear ({\it i.e.}\ the right-hand side of the tilt constraint) and the
metric perturbation vanish in such modes at linear order.

The Friedmann and Raychaudhuri equations
(\ref{eq:friedmann}) and  (\ref{eq:raychaudhuri}) are modified by the
perturbation to $^{3}R$ at first order,
\begin{eqnarray}
\delta(^3Re^{2\alpha}) &=& \sum_{m=1}^5 \mathcal{R}^{(m)} A^{(m)} \nonumber \\
\mathcal{R}^{(m)} &=&  \mathcal{R}_{ab} \beta^{(m)}_{ab} \textrm{,}  \label{eq:R-perturbation}
\end{eqnarray}
with
\begin{equation}
\mathcal{R}_{ab}  = -4n_{ac} n_{cb} + 2 n_{ab} n_{cc} + 12 a_a a_b\textrm{,}
\end{equation}
although in practice most of the $\mathcal{R}^{(m)}$ vanish (Table
\ref{tab:frw-perturb}). The resulting backreaction on the evolution
equation (\ref{eq:evol-in-beta}) is second order and is therefore not
considered.

\subsection{Gauge modes}
\label{sec:gaugemodes}

The perturbed metric in equation~(\ref{eq:perturbmetric}) is not fully
gauge-fixed by the requirements we have outlined (that $|\beta_{ab}|$
be small; that the commutators of the $\vec{e}_a$ are canonical; and
that $[\vec{n},\vec{e}_a]=0$).  There are two residual gauge freedoms:
a global time shift in the labelling of the homogeneous hypersurfaces,
$t \rightarrow t + \delta t$, and a constant linear transformation of
the $\vec{e}_a$ that preserves the canonical form of the structure
constants.  Applying these gauge freedoms to the background FRW metric
generates apparent perturbations that are actually gauge artefacts.

Small time shifts do not affect $\beta_{ab}$ at first order, but
global reparametrizations $\vec{e}_a \rightarrow T_{ab} \vec{e}_b$
generally do. (The latter transformations correspond to
time-independent automorphisms in the group-theoretic terminology.)
Under reparametrization, the structure constants transform
tensorially; hence $a_a$ and $n^{ab}$ transform as components of a
covector and pseudo-tensor respectively:
\begin{eqnarray}
a_a &\rightarrow  T_{ab} a_b \\
n^{ab} &\rightarrow \mathrm{det}\, (T) (T^{-1})_{ca} (T^{-1})_{db}  n^{cd} \, .
\end{eqnarray}
Here, we are concerned with reparametrizations that maintain the
canonical form of $a_a$ and $n^{ab}$.  For such a $T_{ab}$ close to
the identity, we can write $T_{ab} = \delta_{ab} + t_{ab}$ where
$t_{ab} = \mathcal{O}(\epsilon)$. This transformation generates a
non-zero, constant $\beta_{ab} = t_{\langle a b \rangle}$ to first
order in $\epsilon$.  To see how these gauge modes relate to the
eigenmodes in Table~\ref{tab:frw-perturb}, note that any mode with
$\mathcal{S}=0$ admits a solution of equation~(\ref{eq:eigenmode})
giving constant $\beta$ (hence vanishing shear) and, by construction,
the anisotropic curvature is vanishing. Such a model is therefore
isotropic, hence FRW, and the ``perturbation'' $\beta_{ab}$ is pure
gauge.  Working through the Bianchi types with FRW limits, we can
enumerate the allowed $t_{ab}$ that preserve the canonical structure
constants and hence determine the gauge modes. For example, in Type V
the allowed $t_{ab}$ have $t_{a1}=0$ but such transformations can
generate an arbitrary constant (STF) $\beta_{ab}$. This is consistent
with the findings in Table~\ref{tab:frw-perturb} that all $\beta_{ab}$
have $\mathcal{S}=0$ in Type-V models. For the other four types, we
similarly find that the gauge modes correspond to linear combinations
of the eigenmodes in Table~\ref{tab:frw-perturb} with $\mathcal{S}=0$.
There are no gauge modes in Type IX.

The scale-factor $e^{\alpha}$ also generally changes under
reparametrization of $\vec{e}_a$; to linear order, $e^{\alpha}
\rightarrow e^{\alpha}[1+t_{aa}/3]$.  In non-flat models, the change
in scale-factor under a gauge transformation is compensated by a
non-zero $\mathcal{R}_{ab} \beta_{ab}$ to preserve the intrinsic
curvature ${}^{3}R$ of the homogeneous hypersurfaces. For the gauge
modes in Types V and VII$_h$, $t_{aa} = -3 \beta_{11}$ and, since
${}^3R$ is unchanged, we require $\mathcal{R}_{ab} \beta_{ab} = 12
\beta_{11}$ (Type V) or $\mathcal{R}_{ab} \beta_{ab} = 12 h
\beta_{11}$ (Type V). These are consistent with the values
$\mathcal{R}=24$ and $24h$ reported in Table~\ref{tab:frw-perturb} for
those eigenmodes with gauge-mode counterparts and $\mathcal{R}$
non-zero.

\subsection{Classification of modes}\label{sec:classification}

We can classify modes according to their transformation under suitable
rotations of the basis vector fields. Under $\vec{e}_a \rightarrow
R_{ab} \vec{e}_b$, with $R_{ab} R_{ac} = \delta_{bc}$, $\beta_{ab}$
transforms tensorially, {\it i.e.}\ $\beta_{ab} \rightarrow R_{ac}
\beta_{cd} R_{bd}$, thus preserving $\beta_{ab} = 0$ in isotropic
models.  Any mode is either invariant under such a transformation, or
transforms into an equivalent mode.

For Types I and IX, the structure constants are invariant under any
rotation. This means that all modes can be made from a single basis
mode under a combination of rotations and linear superpositions; hence
all modes must have the same dynamical behaviour.  However in Types
VII$_0$, V and VII$_h$ the structure constants possess only a residual
$U(1)$ symmetry corresponding to rotation about $\vec{e}_1$. In
Table~\ref{tab:frw-perturb}, the $s$ modes are invariant under $U(1)$,
whereas $v_1$ and $v_2$ are a pair of modes which transform into each
other under a spin-1 representation. In the case of $t_1$ and $t_2$ for
models other than VII$_h$, the modes transform into each other under a
spin-2 representation. For the VII$_h$ spaces, the diagonalization
procedure applied to the anisotropic curvature
introduces complex elements, describing left- and right-circularly-polarized
gravitational wave modes which do not mix but
individually transform according to a complex spin-2 representation.

Our classification of the Type-V and VII modes into $s$, $v$ and $t$ classes
is analogous to the Fourier-based classification of linear
perturbations about flat FRW models into scalar, vector and tensor
perturbations, whereby scalar, vector and tensor modes transform as
spin-0, 1 and 2 representations under $U(1)$ rotations about the
Fourier wavevector. More generally, cosmological perturbations over
maximally-symmetric 3-spaces can be decomposed into scalar, vector and
tensor modes following~\cite{1990CQGra...7.1169S}.  Adopting the
synchronous gauge, the perturbed line element is
\begin{equation}
  \dd s^2 = - \dd t^2 + a^2(t)[(1+2\phi)\gamma_{ij} + h_{ij}] \dd x^i \dd x^j
\end{equation}
for a background FRW spacetime with spatial metric $a^2(t)\gamma_{ij}$
in some coordinate chart $\{t, x^i\}$. The trace-free $h_{ij}$ decomposes as
\begin{equation}
h_{ij} = (\nabla_i \nabla_j - \gamma_{ij} \nabla^k \nabla_k /3)
\mathcal{E} + 2 \nabla_{(i} B_{j)} + W_{ij} ,
\end{equation}
where $\nabla_i$ is the spatial covariant derivative compatible with
$\gamma_{ij}$, $\mathcal{E}$ is a scalar field, $B_i$ is a solenoidal 3-vector field
and $W_{ij}$ is a STF and transverse 3-tensor field.
The perturbations $\phi$ and $\mathcal{E}$ describe scalar perturbations,
$B_i$ vector perturbations and $W_{ij}$ tensor perturbations.
In the closed case, the decomposition into scalar, vector and tensor
parts is unique but in the non-compact case this is only true if the
perturbations satisfy appropriate asymptotic decay conditions.

Our $t$-mode metric perturbations can all be shown to be transverse and
so can be interpreted as tensor perturbations. However this
interpretation is generally not unique since the required asymptotic
decay conditions are not met here. Similarly, our $v$-mode metric
perturbations can be interpreted as vector modes, {\it i.e.} they can be
written as $\nabla_{(i} B_{j)}$ for a solenoidal vector $B_i$. For the
$v$-modes, it is always possible to take $B_i$ to be homogeneous;
explicit constructions are given in Table \ref{tab:vector-mode-generators}.

\begin{table}
\begin{center}
\begin{tabular}{rl}
Mode & Vector generator $\vec{B}$ \\
\hline
V\,$v_1$ & $\vec{e}_2$ \\
$v_2$ & $\vec{e}_3$ \\
\hline
VII\,$v_1$ & $\left(\sqrt{h} \vec{e}_2 -\vec{e}_3\right)/\left(h+1\right)$ \\
$v_2$ & $\left(\vec{e}_2 + \sqrt{h}\vec{e}_3\right)/\left(h+1\right)$ \\
\hline
\end{tabular}
\end{center}
\caption{The generators for the vector modes in Types V, VII$_0$ and VII$_h$
  relative to the invariant $\vec{e}_i$ basis. For the Type-VII$_0$ generators,
$h$ should be set to zero.
Note that in all types displayed,
$\vec{e}_2$ and $\vec{e}_3$ are solenoidal. 
The $\vec{e}_1$ direction does
  not feature because it is either Killing (Type VII$_0$), in which case
$\nabla_{(i}B_{j)}$ vanishes, or not solenoidal (Types V and VII$_h$).}
\label{tab:vector-mode-generators}
\end{table}

The metric perturbation from $\beta_{ab}$ (see equation~\ref{eq:metricpert})
can be shown to be transverse for all Type-IX modes and, since the
spatial sections are compact, these are uniquely to be interpreted as tensor
perturbations
(gravitational waves \cite{1991PhRvD..44.2356K}). One can also show
(Section \ref{sec:coord-expr-modes}) that the invariant fields are
Killing, which explains the lack of any homogeneous vector
perturbations.

 The ambiguity in the scalar, vector and tensor decomposition in
the non-compact case is simply illustrated in 
Type I since the
background is flat space. All modes are transverse but may
nevertheless be expressed in terms of vector modes: consider, in a
coordinate basis $\{x,y,z\}$, the Type-I mode with only
$\beta_{12}=\beta_{21} = 1$ non-zero.
One may decompose
$\beta_{ij} = \partial_{(i} V_{j)}$ with $\vec{V}= (x \partial_y +
y \partial_x)$. The vector $\vec{V}$ is solenoidal ($\nabla \cdot
\vec{V}=0$), but also curl free ($\nabla \times \vec{V}=0$), so that
it too can be further decomposed into the scalar $\vec{V} = \nabla
\phi$, $\phi = xy$. Similar statements can be made for any Type-I
mode. These decompositions are possible in spite of standard results
because $\beta_{ij}$ does not decay towards infinity.

\subsection{Metric perturbation in coordinate bases}

The expressions in Table \ref{tab:frw-perturb} can be expanded
relative to a coordinate basis by using the explicit expressions for
the triad of invariant vector fields given in Section
\ref{sec:coord-expr-modes} and \ref{sec:altern-coord-expr}.

The Type-I case is trivial, since the invariant vectors are simply the
coordinate differentials, $\vec{e}_i = \partial_i$. We therefore
consider, for illustration, the next simplest case, that of the
VII$_0$ modes on a flat FRW background.  In this case, expressions
(\ref{eq:basis-vii0}) can be used to generate an anisotropic synchronous-gauge
metric perturbation,
\begin{equation}
\delta g_{\mu\nu} = 2 e^{2\alpha} \beta_{ab} (e^a)_\mu (e^b)_\nu ,
\label{eq:metricpert}
\end{equation}
for each decoupled linear Bianchi mode.

For example, the VII$_0$ $v_1$ perturbation has an anisotropic
correction to the metric with coordinate-basis components $(\delta
g_{12}, \delta g_{13})$ proportional to $(\cos x^1, \sin x^1)$. For
the VII$_0$ $s$ perturbation, the anisotropic metric perturbation has
components $(\delta g_{11}, \delta g_{22}, \delta g_{33})$
proportional to $(2,-1,-1)$.  In this case, we can regard the mode
also as a Type-I perturbation, {\it i.e.}\ the metric is invariant under the
action of both the Type-I and VII$_{0}$ groups and the perturbed space
is locally rotationally symmetric, having four KVFs ($\{\vec{T}_i\}$
and $\vec{R}_1$). Finally, the VII$_0$ $t_1$ perturbation produces an
anisotropic metric perturbation with $\delta g_{22} = - \delta g_{33}
\propto \cos 2x^1$ and $\delta g_{23} \propto \sin 2x^1$. Physically,
this is a standing wave resulting from the superposition of two
counter-propagating left-circularly polarized gravitational waves with
wavenumber~2.

\section{Relation to non-linear solutions}\label{sec:stab-line-relat}

Compared with previous approximate approaches to analysing the
behaviour of near-isotropic Bianchi models, the advantage in the
description above is its use of the physical anisotropic curvature and
shear variables as natural linearisation coordinates.  The
linearisation applies when anisotropic shear, stress and curvature are
all small; this is expected to be suitable for describing any residual
anisotropy in the Universe today but may break down at arbitrarily
early or late times. Therefore we now place our evolution equations in
the wider context of present knowledge of the properties of the exact
solutions.

\subsection{Early-time attractors}

For a given shear at the present epoch, it is usually supposed that
anisotropies were larger at sufficiently early times when the FRW
scale-factor was smaller; ultimately as $\alpha \to -\infty$ one
expects to enter a non-linear regime in which at least one of the
conditions (\ref{eq:linear}) breaks down. While this expectation is
broadly confirmed, we will discuss below that there are a few
exceptions.

The supposition of a non-linear phase arises from the early-time
attractor for open and flat models \cite{1973JETP...37..739D}, which
is a Kasner shear-dominated approach to the singularity. In that case,
all isotropic and anisotropic curvature lengthscales become
super-horizon and the Universe is indistinguishable from a Type-I
model with shear diverging as $(1+z)^3$ for $z \to \infty$. Our
description does not explicitly resolve the Kasner vacuum behaviour
because we have neglected $\sigma^2$ terms in the Friedmann constraint
(\ref{eq:friedmann}) and Raychaudhuri equation
(\ref{eq:raychaudhuri}).  The overall dynamics of such a model may be
understood by matching an early Kasner phase onto our late-time linear
description.

However it need not be the case that lengthscales of interest are
super-horizon at early times: for instance in Misner's closed
`mixmaster' (Type-IX) Universes \protect\cite{1969PhRvL..22.1071M},
quasi-Kasner phases are interspersed with `bounces' which reverse the
sign of the shear. These bounces occur when the universe becomes
sufficiently small along two of its principal axes that the curvature
re-enters the horizon. (Pancaking universes, in which only one axis
becomes short, collapse to a singularity without a bounce; this
corresponds to travelling along the $\beta_+$ direction in Misner's
terminology.)

In fact even in open and flat models, the Kasner attractor may be
avoided: Lukash \cite{1976NCimB..35..268L} showed that in a VII$_0$ or
VII$_h$ model the initial conditions may be chosen so that the
anisotropy in gravitational waves is hidden in super-horizon curvature
at early times. In our notation this corresponds to setting
$A^{(t)\prime}=0$, $A^{(t)}=A_0$ at an initial singularity. Inspecting
equation (\ref{eq:eigenmode-evol}) and its eigenvalues (Table
\ref{tab:frw-perturb}) shows that, with such initial conditions, shear
is generated in the tensor modes only once the mode enters the horizon
({\it i.e.} $\eta\sqrt{\mathcal{S}} > 1$), in agreement with Lukash's
description.  The dynamical behaviour is then correctly modelled all
the way back to the initial singularity using our linear
approximation. 

The existence of these modes has been highlighted by various authors,
for instance Collins and Hawking (who refer to the existence of a
`growing mode' due to the behaviour at very early times
\cite{1973MNRAS.162..307C}) and Barrow and Sonoda
(Ref. \cite{1986PhR...139....1B}, Section 4.9; the regular $t$ modes
and decaying $s$ modes are seen, but not the $v$ modes because the
tilt is set to zero).  In our own work we have previously
\cite{2009PhRvD..79j3518P} denoted these modes $t_1$ (distinct from
the shear-diverging $t_0$). Note that, contrary to the notation in
that work, the present paper's $t_1$ and $t_2$ modes are rotations (or
conjugates) of each other, both able to give regular high-redshift
behaviour if the initial data specifies $A'=0$.

Similar tuning of the initial conditions can be applied to produce
closed Type-IX models with a non-oscillatory approach to the initial
singularity, despite their generic mixmaster behaviour described
above. As with Type VII, the existence of regular Type-IX early-time
solutions can be understood through the decomposition into an exact
background with super-horizon gravitational waves; this separation has
been further explored in Refs.  \cite{1975JETP...42..943G} and
\cite{1991PhRvD..44.2356K}.

\subsection{Late-time attractors}

We have considered above the `early-time' behaviour, by which we mean
the behaviour when isotropic and anisotropic curvature scales are
super-horizon. We will now turn to the `late-time' behaviour of models
after the characteristic scales are sub-horizon. We should note,
however, that the real Universe could be in either of these states --
{\it i.e.}\ there is no guarantee that we have now entered a `late time' in
terms of the dynamical evolution of anisotropies in the Universe (if
such anisotropies indeed exist).

Near-isotropic Type-I, V and IX models are relatively uninteresting at
late times. Type-I and V models which are sufficiently close to
isotropy will always tend to full isotropy at sufficiently late times
\cite{1973MNRAS.162..307C}. This is reflected in the trivial behaviour
of the linear dynamical modes, all of which have $\mathcal{S}=0$
(Table \ref{tab:frw-perturb}) and
so have shear that decays as $(1+z)^3$. Since Type-IX
models recollapse at late times, they do not achieve a late-time limit
but instead re-enter a time-reversed `early-time' phase.

Of more interest are the Type-VII models since, according to Collins
and Hawking \cite{1973MNRAS.162..307C}, the subset which ultimately
isotropize is of measure zero in the initial conditions. This
conclusion is reinforced by the more recent numerical analysis of
Wainwright \cite{1998CQGra..15..331W}; the late-time behaviour of the
Type-VII models therefore warrants closer inspection.

The quantities which define isotropization for each mode are the
expansion-normalized shear ($\Sigma^{(m)} =
A^{(m)\prime}/\mathcal{H}$) and curvature ($
\mathcal{S}^{(m)}A^{(m)}/\mathcal{H}^2$). Isotropization occurs if
both of these tend to zero\footnote{As noted by various authors, this
  restriction is stronger than required by the observational
  constraints on isotropy; see the last paragraph of Section
  \ref{sec:other-bianchi-types}.} as $t \to \infty$; conversely the
universe becomes arbitrarily anisotropic (and our linear approximation
breaks down) if either quantity diverges.  We will first consider the
behaviour of flat VII$_0$ models, followed by the open VII$_h$ models.
In both cases only the $t$ modes are of interest, since all other
modes have no interaction with the anisotropic curvature and the shear
decays as $(1+z)^3$.

For $t$ modes of type VII$_0$ at late times, equation
(\ref{eq:eigenmode-evol}) shows (for $\gamma=p/\rho+1 >2/3$) that
$\Sigma^{(t)} \propto e^{2i\eta} e^{\alpha(3\gamma/2-2)}$. In the case
of radiation domination, $\gamma=4/3$, the expansion-normalized shear
thus undergoes constant-amplitude oscillations; at second order the
amplitude has been shown to decay logarithmically
\cite{1973JETP...37..739D}. Such slow decay has been emphasized by
Barrow \cite{1995PhRvD..51.3113B,1997PhRvD..55.7451B} in the context
of nucleosynthesis constraints, but it is important to realize that it
applies only to the {\it late-time} expansion in which
$\sqrt{\mathcal{S}^{(m)}}\eta \gg 1$.  This will be attained during
radiation domination only if
the current physical length scale of the spiralling is smaller
than $\Omega_{R}^{1/2}r_H/\Omega_{M}$, where $r_H$ is the present-day
Hubble length and $\Omega_R$ and $\Omega_M$ are the current radiation and matter
density parameters.
If this condition is not satisfied, the shear and
anisotropic curvature of the tensor modes were essentially decoupled
during radiation domination because the scale-length was
super-horizon.

In Type-VII$_h$ models the conformal Hubble parameter $\mathcal{H}$ is
constant but non-vanish\-ing at late times once isotropic curvature
dominates. (The following considerations therefore do not apply to
models with dark energy.)  In our linear approximation the solutions
for the $t$ modes in the curvature-domination limit are
\begin{equation}
A(\eta) = \exp \left\{-\mathcal{H} \left[1 \pm \sqrt{1-\mathcal{S}^{(m)}/\mathcal{H}^2}\right]\eta\right\} ,
\end{equation}
with $\mathcal{H}=+\sqrt{h}$ and $\mathcal{S}^{(m)}=4
(1-i\sqrt{h})$. The amplitude of the oscillations is determined by $1
\pm \mathrm{Re}\, \sqrt{1-\mathcal{S}^{(m)}/\mathcal{H}}$, which is
equal to $1\pm 1$ for all $h$. Consequently the amplitude of the
expansion-normalized shear oscillation is either constant or decays as
$e^{-2\alpha}$. The `frozen-in' constant-amplitude mode was previously
noted in the analysis of Doroshkevich {\it et al.}
\cite{1973JETP...37..739D}.

\subsection{Implications of nucleosynthesis in the observed universe}

The above considerations have implications for constructing a Bianchi
model which is compatible with big-bang nucleosynthesis (BBN)
constraints. Any remaining difficulties in fully reconciling observed
abundances with theory based on the fiducial FRW picture
(e.g. Ref.~\cite{2006IJMPE..15....1S}) are relatively minor; BBN thus
constitutes significant evidence against any model which produces an
expansion history significantly different from FRW at $T \simeq 10^9
\, \mathrm{K}$.  In fully non-linear relativistic evolution, shear
enhances deceleration, modifying the Friedmann (\ref{eq:friedmann})
and Raychaudhuri (\ref{eq:raychaudhuri}) equations so that
\begin{eqnarray}
\rho &=& 3 H^2 - \sigma^2 + \tensor[^{3}]{R}{}/2 - \Lambda\\
\dot{H} &=& -H^2 - \frac{1}{6}(\rho + 3 p) + \frac{\Lambda}{3} - \frac{2}{3} \sigma^2 \textrm{,}
\end{eqnarray}
where $\sigma^2 \equiv
\sigma_{\hat{a}\hat{b}}\sigma^{\hat{a}\hat{b}}/2$.  For this
reason, significant shear at high redshifts is prohibited by our
knowledge of the chemical composition of the Universe. Since for most
linearized modes $\sigma \propto (1+z)^3$ (this relation also holding
exactly for the non-linear Type-I models), even weak shear constraints
at high redshift translate into stringent upper bounds at the present
day. Accordingly, for many models, BBN is known to constitute a strong
limit (competitive with or even stronger than microwave background
bounds -- see \cite{1976MNRAS.175..359B}). However, exceptions arise
in certain models which, in terms of our description, either have
regularized initial conditions (the $t$ modes in which shear
anisotropy vanishes and anisotropic curvature scales are super-horizon
at high redshift), or much slower than $(1+z)^3$ decay at late times
(the logarithmically decaying modes for very small spiral scales). In
such circumstances, the BBN constraints offer far weaker limits on
present day shear; it is these models, therefore, which should be of
most interest to CMB phenomenologists \cite{2009PhRvD..79j3518P}.

\section{Other Bianchi types}\label{sec:other-bianchi-types}

We conclude with some comments on the Bianchi models which do not
possess a strictly isotropic limit (Types II, IV, VI$_0$, VI$_h$ and
VIII), but nonetheless include models which are close to isotropy. By
definition these models can still be thought of (inside the horizon
for a limited epoch) as perturbations on a maximally-symmetric FRW
background, but the symmetries of the perturbations will no longer
exactly correspond to Killing fields of the background
maximally-symmetric models.

For near-isotropy on any single timeslice, the expansion-normalized
shear $\sigma_{\hat{a}\hat{b}}/H$ and anisotropic curvature $^3
S_{\hat{a}\hat{b}}/H^2$ must be $\mathcal{O}(\epsilon)$. Suppose we
now take a general Bianchi model, not necessarily with a strict FRW
limit, which on some timeslice satisfies these requirements. On this
timeslice we may, without producing a physical change, reparametrize
the triad of invariant vectors so that $\beta_{ab}$ is zero. This may
bring the commutation constants $C_{ab}^c$ into a non-canonical form,
but we can at least rotate the basis vectors so that $n_{ab}$ is
diagonal and $a_i=0$ for $i>1$. We then form a tetrad with the
timelike normal and extend to nearby timeslices using the same
time-invariant prescription described in Section
\ref{sec:enum-line-bianchi}.  At times sufficiently near our original
chosen timeslice, the conditions described by (\ref{eq:linear}) will
apply so that our perturbative expansion of the dynamics is valid.

The physical quantity $^3 S_{\hat{a}\hat{b}}/H^2$ expressed on an
orthonormal tetrad remains invariant (up to rotations) under the above
transformations. Since on the preferred timeslice we have set
$\beta_{ab}=0$, only the zero-order part of expression
(\ref{eq:iso-curv-expand}) for this quantity contributes.  It follows
that the structure constants of the time-invariant tetrad must be
equal to those of a `nearby' isotropic Bianchi type (for which the
zero-order contribution to $^3S_{\hat{a}\hat{b}}$ would vanish), plus
a small perturbation. In such a case of near isotropy, all those Bianchi types that do
not have a strict isotropic limit must fall into one of the following three
possibilities.

\begin{enumerate}
\item  All of the structure constants are small in the sense that
\begin{equation}
  e^{-2\alpha} |C^2/H^2| = \mathcal{O}(\epsilon)\textrm{.}
\end{equation}
The nearby isotropic case is Bianchi Type I, and the actual universe could be any
Bianchi type. Because the structure constants are small, the
anisotropic part of the CMB \cite{2007MNRAS.380.1387P} appears as a
pure quadrupole at first order provided the near isotropy holds between recombination
and the present epoch. The dynamics are modified such that
\begin{equation}
\beta_{\hat{a}\hat{b}}'' + 2 \mathcal{H} \beta'_{\hat{a}\hat{b}} + \mathcal{S}_{\hat{a}\hat{b}} = 0,
\end{equation}
where $\mathcal{S}_{\hat{a}\hat{b}} = {}^3 S_{\hat{a}\hat{b}} e^{2 \alpha}$
is time independent at first order.

In those models with an exact FRW limit, we are describing a situation
where the shear is small and the isotropic curvature scale and any
twist scale are small compared to the Hubble radius.  In other models,
while the linearization applies, we can say the Universe is in a flat
quasi-Friedmann epoch, in agreement with the terminology of
Ref. \cite{wainwright1997dsc} (see Section
15.2.1). $\mathcal{S}_{\hat{a}\hat{b}}$ can be made arbitrarily small
by appropriately scaling the structure constants. The epoch of
apparent isotropy can thus be made arbitrarily long in any Bianchi
type, although this implies fine-tuning of the initial conditions.

\item $a_1=1$, $n_1=0$; $n_2$ and $n_3$ are both small in the sense that
\begin{equation}
e^{-2\alpha} a_1 | n_i/H^2 | \le \mathcal{O}(\epsilon)\label{eq:near-v-criterion}
\end{equation}
The nearby isotropic case is Type V. One possibility for the actual
Bianchi model under analysis is Type VII$_h$ ($\mathrm{sign}\,
n_2=\mathrm{sign}\,n_3$). Building a rescaled triad shows that this
captures VII$_h$ models in the limit where the
twist scale is large compared to the Hubble radius.

The other possible types which can
achieve near-type-V isotropy are Type VI$_h$ ($\mathrm{sign}\,
n_2=-\mathrm{sign}\,n_3$) or IV (one of $n_2$ or $n_3$ vanishes).
In all types, the expression for the anisotropic curvature reads
\begin{equation}
^3S_{\hat{a}\hat{b}} e^{2\alpha}= \mathcal{S}_{cd\hat{a}\hat{b}}^{\mathrm{V}} \beta_{cd} + (n_2-n_3)\beta^{t_2}_{\hat{a}\hat{b}} = (n_2-n_3)\beta^{t_2}_{\hat{a}\hat{b}} ,
\end{equation}
which drives the corresponding gravitational wave term in Table
\ref{tab:frw-perturb} according to

\begin{equation}
A^{(t_2) \prime \prime} + 2 \mathcal{H} A^{(t_2) \prime}  = -(n_2-n_3) \textrm{ . }
\end{equation}
The universe can then be said to be in an open quasi-Friedmann epoch.

\item $|n_2-1|$ and $|n_3-1|$ are small; $a_1=0$; $n_1$ is small such that
\begin{equation}
e^{-2\alpha} |n_2^2 -n_3^2|/H^2 = \mathcal{O}(\epsilon) , \qquad
e^{-2\alpha} |n_1(n_2+n_3)|/H^2 = \mathcal{O}(\epsilon) .\label{eq:near-VII0-restrictions}
\end{equation}
  The nearby isotropic case
  is Type VII$_0$ and the actual Bianchi model under analysis is Type
  VIII ($n_1<0$) or IX ($n_1>0$).

  Building a triad with canonical Type-IX commutation relations from a
  near-VII$_0$ model will produce a decomposition with highly
  anisotropic metric ($|\beta_{ab}|$ large), violating the original
  linearisation conditions (\ref{eq:linear}).  The near-VII$_0$ models
  therefore form a distinct isotropic limit which only the alternative
  analysis of this section will correctly capture.

  For all near-VII$_0$ types, the modified curvature expression reads
\begin{equation}
^3S_{\hat{a}\hat{b}} = \mathcal{S}^{\mathrm{VII}_0}_{cd\hat{a}\hat{b}} \beta_{cd} - \frac{n_1}{3} \beta^{s}_{\hat{a}\hat{b}} ,
\end{equation}
which drives the scalar mode, {\it i.e.}
\begin{equation}
A^{(s) \prime \prime} + 2 \mathcal{H} A^{(s) \prime}  = \frac{1}{3}n_1 \textrm{ . }\label{eq:driven-shear-near-vii0}
\end{equation}
(Note that if only this driven scalar mode is present, the values of
$n_2$ and $n_3$ are unobservable since $\beta_{ab}$ has a $U(1)$ symmetry about
the $\vec{e}_1$ direction; the physical situation is then
identical to a near-Type-I model with small $n_1$.)
\end{enumerate}

We have already commented that in the near-Type-I case, the
anisotropic-sourced part of the CMB remains a pure quadrupole if the
linearisation holds all the way back to recombination.  Given the
constraints on flatness in today's Universe, we also know that the
near-Type-V case could be seen only in a near-flat limit which,
likewise, would involve only a quadrupole contribution to the CMB.  On
the other hand the near-VII$_0$ models (case iii) give rise to
more interesting insights, as described below.

\subsection{Near VII$_0$: spiral patterns in Type-IX models}

Case (iii) of the preceding text shows that spiral patterns in the CMB
can arise in Type-IX models.  Specifically, by perturbing $n_1$ about
a VII$_0$ background with $t$ or $v$ modes on some initial
hypersurface, one will generate a closed Type-IX universe which has
observable spiral structure in the CMB. Note that, because it is the
{\it difference} $|n_2^2-n_3^2|^{1/2}$ which must be small compared to
the inverse comoving horizon scale $H e^{\alpha}$, the VII$_0$ spiral
scale (set by the magnitude of $e^{\alpha}/n_2$ or $e^{\alpha}/n_3$)
can be sub-horizon.  This is a surprising conclusion: Grishchuk showed
\cite{1975JETP...42..943G} that all Type-IX models can be decomposed
into an exact FRW background and maximal-wavelength gravitational
waves -- the origin, in this picture, of an apparently independent
spiral scale is at first unclear.

However, one can confirm the spiralling even in the Grishchuk
decomposition by, for instance, transforming a model obeying
restrictions (\ref{eq:near-VII0-restrictions}) into a frame in which
the structure constants are canonical. In this ``Grishchuk frame'',
$\beta_{ab}$ is nonlinear even though the shear and anisotropic
curvature remain small while the mode is on super-Hubble scales.  The
exact geodesic equation (e.g. equation 23 in
Ref. \cite{2007MNRAS.380.1387P}) can be applied; because the
Grishchuk-frame components of $e^{2\beta}$ are far from the identity
matrix, the geodesic behaviour is necessarily altered from the
canonical isotropic Type-IX model. Once the resulting geodesic is
expressed relative to an orthonormal group-invariant triad, one sees
multiple spirals per Hubble length, in exact agreement with the
effects derived from the original near-VII$_0$ frame. In other words,
the spiral scale information is correctly encoded in the non-linear
gravitational waves of the Grishchuk picture.

Before leaving this topic, we should note that spiralling in Type-IX
has recently been discovered independently by a quite different
approach \cite{Lasenby2010}.


\subsection{Isotropic curvature and anisotropy}

For the near-V and near-VII$_0$ models (cases ii and iii), the horizon
re-entry of isotropic curvature will unavoidably trigger the
appearance of anisotropy in the universe, because the anisotropic- and
isotropic-curvature scales are geometrically linked. For instance,
consider the Type-IX models which are passing through a near-VII$_0$
phase. To first order, the Friedmann constraint (\ref{eq:friedmann})
reads
\begin{equation}
3 \mathcal{H}^2 = \rho a^2 -n_1\textrm{.}
\end{equation}
Consequently a detection of the universe's closure is possible once
the expansion has slowed sufficiently for $n_1/\mathcal{H}^2$ to
exceed some small threshold. A similar criterion applies to the
detectability of the driven scalar shear mode
(\ref{eq:driven-shear-near-vii0}), if its magnitude is initially close
to zero.  So at the same time as the isotropic curvature enters the
horizon, the scalar-mode driving term will become important.

In the linear approximation, there are no late-time solutions with
small anisotropies in any of the models introduced in this section,
unless $\gamma<2/3$.  At sufficiently late times, the near-isotropy
will necessarily break down for these classes of models, unless some
form of dark energy is dominant or the expansion is reversed. In terms
of the exact Bianchi solutions for $\gamma>2/3$, the near-isotropic
solutions form saddle points since they have both stable directions
(decaying shear modes) and unstable directions (growing modes, those
which are forced by the small 3-curvature).

It has previously been emphasized \cite{1973JETP...37..739D}, however,
that the existence of long-lived quasi-FRW solutions in all Bianchi
types is all that is required to make those Bianchi types of
observational relevance; even without dark energy there would be no
observational need for the isotropization to be asymptotic as $t \to
\infty$. All this makes it impossible to rule out the relevance of any
given Bianchi type to the observed universe, although the
observational near-flatness places partial constraints on the scale of
the Bianchi geometry relative to today's horizon.

\section{Summary}\label{sec:summary}

We have investigated in some detail the geometrical origin and
time-dependent behaviour of Bianchi perturbations on
maximally-symmetric ({\it i.e.} FRW) backgrounds. Many results obtained
relate closely to previously-known properties, but to our knowledge
this work is the first to consider the various models in a unified way
using physically transparent variables throughout.

By diagonalising the linearised Einstein equations, we have enumerated
the linear modes in such models, and have shown each to be a scalar,
vector or tensor mode on a suitable background. It is worth noting
that there is nothing particularly special about these modes to
suggest they should be excited by physical processes (with the
possible exception of Type IX, which may be written as maximal-wavelength gravitational waves on a closed background
\cite{1991PhRvD..44.2356K}). In particular, Type-V, VII$_0$ and
VII$_h$ modes have symmetries which bear no relation to geodesics on
the background manifold. While making the models interesting to CMB
phenomenologists \cite{2005ApJ...629L...1J,2007MNRAS.380.1387P},
exactly this non-geodesic property makes them rather hard to motivate
as candidates for ``natural'' initial conditions. On the other hand,
the absence of any other way to transport anisotropic quantities
around open spaces means that the Bianchi models must be taken
seriously by any cosmologist who does not on physical grounds rule out
the possibility of either an open or an anisotropic Universe.

We closed by inspecting those Bianchi types which do not possess a
strictly isotropic limit, concluding that all Bianchi types can appear
isotropic and flat for an arbitrary period of physical time.  We will
presumably never be able to rule out observationally the possibility
that we live in a Universe which is highly anisotropic on today's
super-horizon scales. In particular, and in agreement with Wald's
theorem~\cite{1983PhRvD..28.2118W}, the behaviour of all universes is
seen to become increasingly isotropic if the late time is dominated by
dark energy.

The first-order evolution equations are extremely simple, and suitable
for direct incorporation into CMB calculations; we have recently
published temperature and polarization maps \cite{2009PhRvD..79j3518P}
by combining the results in the present work with our earlier
derivation of the Boltzmann hierarchy for all nearly-isotropic Bianchi
models \cite{2007MNRAS.380.1387P}, excluding the types explained in
Section \ref{sec:other-bianchi-types} (the latter models would yield
either a pure quadrupole or similar morphology to Type-V or Type-VII$_0$ maps).  Of course, although we tried to show (in Section
\ref{sec:stab-line-relat}) that our analysis does offer a different
way of thinking about many previously known results, only a shadow of
the complex behaviour which makes these models dynamically interesting
\cite{wainwright1997dsc} is captured in detail.

\ack

AP is supported by a Research Fellowship at Emmanuel College,
Cambridge.  We thank John Barrow, Steven Gratton, Sigbj\o rn Hervik,
Anthony Lasenby and Matthew Johnson for useful discussions.

\section*{References}
\bibliographystyle{unsrt} \bibliography{../refs.bib}

\appendix

\section{Alternative coordinate expressions for Killing fields and invariant vectors}
\label{sec:altern-coord-expr}

In Section \ref{sec:coord-expr-modes}, we gave expressions in terms of
coordinate charts for the Killing fields of flat, open and closed
maximally-symmetric spaces. We related these to the subset of fields
which generate the Bianchi symmetries and listed the invariant fields
for those cases. In this appendix we give, for completeness, these
results in some alternative coordinate systems in the open and closed
cases. As previously emphasized, these can be used to generate explicit
expressions for the metric perturbations generated by each of the
modes in Section \ref{sec:enum-line-bianchi}.

\subsection{Open space}
\label{subsec:open_app}

In order to express the results of Section \ref{sec:open-space} in
terms of familiar coordinates, we define $r$, $\theta$, $\phi$ by the
following:

\begin{eqnarray}
a^0 = \cosh r  & a^1 = \sinh r \cos \theta \nonumber \\
a^2 = \sinh r \sin \theta \cos \phi \hspace{0.5cm} &  a^3 = \sinh r \sin \theta \sin \phi \textrm{.}
\end{eqnarray}

\noindent
The metric induced by (\ref{eq:open-ambient}) is:
\begin{equation}
\dd s^2 = \dd r^2 + \sinh^2 r \left(\dd \theta^2 + \sin^2 \theta~ \dd \phi^2\right)\textrm{.}
\end{equation}
With a little algebra, one may then obtain the KVFs in terms of
the coordinate basis ${r,\theta,\phi}$:
\begin{eqnarray}
\hspace{-2cm}\vec{{\bm \xi}}^{\mathrm{V}}_1 = \cos \theta \partial_r - \coth r \sin \theta \partial_{\theta} \nonumber \\
\hspace{-2cm}\vec{{\bm \xi}}^{\mathrm{VII}_h}_1 = -\sqrt{h} \cos \theta \partial_r + \sqrt{h} \coth r \sin \theta  \partial_{\theta} + \partial_{\phi} \nonumber \\
\hspace{-2cm}\vec{{\bm \xi}}^{\mathrm{V}}_2 = \sin \theta \cos \phi \partial_r + \left(\cos \theta \coth r  -1 \right) \cos \phi~ \partial_{\theta} + \left(\cot \theta - \coth r \csc \theta \right) \sin \phi~ \partial_{\phi} \nonumber \\
\hspace{-2cm}\vec{{\bm \xi}}^{\mathrm{V}}_3 = \sin \theta \sin \phi \partial_r + \left(\cos \theta \coth r  - 1 \right) \sin \phi~ \partial_{\theta} + \left( \coth r \csc \theta -\cot \theta  \right) \cos \phi~ \partial_{\phi} 
\end{eqnarray}
The ${\bm \xi}_2$ and ${\bm \xi}_3$ in Type VII${}_h$ follow from equation~(\ref{eq:v-becomes-viih}).
Near to $r=0$, the Type-V KVFs reduce to Type I ({\it i.e.}\ translations along
local Cartesian axes) as expected. Note also that the KVFs are well-defined
everywhere despite the coordinate singularity at $r=0$.
The Type-V invariant vectors become
\begin{eqnarray}
\hspace{-2cm}\vec{e}_1^{\mathrm{V}} = \frac{\cos \theta \cosh r - \sinh r}{\cosh r - \cos \theta \sinh r} \partial_r - \frac{\mathop{\mathrm{csch}} r \sin \theta}{\cosh r - \cos \theta \sinh r} \partial_{\theta}  \nonumber \\
\hspace{-2cm}\vec{e}_2^{\mathrm{V}} = \frac{\sin \theta \cos \phi}{\cosh r - \cos \theta \sinh r} \partial_r + \frac{\cos \phi\left(\cos \theta \coth r -1\right)}{\cosh r - \cos \theta \sinh r} \partial_{\theta} - \sin \phi \csc \theta \mathop{\mathrm{csch}} r \partial_{\phi} \nonumber \\
\hspace{-2cm}\vec{e}_3^{\mathrm{V}} = \frac{\sin \theta \sin \phi}{\cosh r - \cos \theta \sinh r} \partial_r + \frac{\sin\phi\left(\cos\theta\coth r-1\right)}
{\cosh r -\cos \theta \sinh r} \partial_{\theta} + \cos \phi \csc \theta \mathop{\mathrm{csch}} r \partial_{\phi} \textrm{ .}
\end{eqnarray}
The invariant basis for Type VII$_h$ follows from equation~(\ref{eq:link-V-to-VIIh-invariant}) with $x=-\ln(\cosh r - \sinh r \cos\theta)$.

An alternative form of the metric follows from foliating an open space
into a series of flat 2-surfaces, resulting in
\begin{equation}
\dd s^2 = \dd x^2 + e^{-2 x}(\dd y^2 + \dd z^2) ,
\end{equation}
where $x$ is as previously defined. The relation between this metric
and the more usual cosmological metric is discussed in Ref.
\cite{1997PhLA..233..169B}. The coordinates are naturally adapted to the
KVFs: starting at $a^0=1$, we reach any point on the open space by
displacing a parameter distance $x$ along ${\bm \xi}^{\mathrm{V}}_1$, followed
by $y$ along ${\bm \xi}^{\mathrm{V}}_2$ and finally $z$ along ${\bm \xi}^{\mathrm{V}}_3$.
We then have
\begin{eqnarray}
a^0 &=& \frac{1}{2} e^{-x} \left(y^2 + z^2\right) + \cosh x \nonumber \\
a^1 &=& \frac{1}{2} e^{-x} \left(y^2 + z^2\right) + \sinh x \nonumber \\
a^2 &=& e^{-x} y \nonumber \\
a^3 &=& e^{-x} z .
\end{eqnarray}
Because they are better adapted to the
Bianchi symmetries, the KVFs and invariant fields look considerably
simpler in these coordinates:
\begin{eqnarray}
\vec{e}_1^{\mathrm{V}} = \partial_x & &\hspace{2cm}\vec{{\bm \xi}}^{\mathrm{V}}_1=  \partial_x + y \partial_y +z \partial_z \nonumber \\
\vec{e}_2^{\mathrm{V}} = e^x \partial_y & &\hspace{2cm}\vec{{\bm \xi}}^{\mathrm{V}}_2 = \vec{{\bm \xi}}^{\mathrm{VII_h}}_2/\sqrt{h} = \partial_y \nonumber \\
\vec{e}_3^{\mathrm{V}} = e^x \partial_z  &&\hspace{2cm} \vec{{\bm \xi}}^{\mathrm{V}}_3 = \vec{{\bm \xi}}^{\mathrm{VII_h}}_3/\sqrt{h} =\partial_z 
\end{eqnarray}
\begin{eqnarray}
\vec{{\bm \xi}}^{\mathrm{VII_h}}_1 = \sqrt{h} \partial_x+(y\sqrt{h}-z) \partial_y + (z\sqrt{h}+y) \partial_z ,
\end{eqnarray}
where the $\vec{e}_i^{\mathrm{VII}_h}$ can be obtained from
(\ref{eq:link-V-to-VIIh-invariant}).

\subsection{Closed space}

Suitable coordinate expressions for the results in Section
\ref{sec:clos-spac} may be expanded from
equation~(\ref{eq:kvf-ix}). We include them here for the usual
spherically-symmetric coordinate chart (for other common systems see
\cite{1991PhRvD..44.2356K}),
\begin{eqnarray}
a^0 = \cos r &  a^1 = \sin r \cos \theta \nonumber \\
a^2 = \sin r \sin \theta \cos \phi \hspace{0.5cm} &  a^3 = \sin r \sin \theta \sin \phi ,
\end{eqnarray}
for which the induced line element is
\begin{eqnarray}
\dd s^2 = \dd r^2 + \sin^2 r \left(\dd \theta^2 + \sin^2 \theta~ \dd \phi^2\right) .
\end{eqnarray}
The KVFs in this chart are 
\begin{eqnarray}
\hspace{-2.5cm}2\vec{{\bm \xi}}_1^{\mathrm{IX}} = -\cos \theta \partial_r + \cot r \sin \theta \partial_{\theta} + \partial_{\phi} \nonumber \\
\hspace{-2.5cm}2\vec{{\bm \xi}}_2^{\mathrm{IX}} = - \cos \phi \sin \theta \partial_r - (\cos \theta \cos \phi \cot r + \sin \phi) \partial_{\theta} + (\cot r \csc \theta \sin \phi - \cos \phi \cot \theta) \partial_{\phi} \nonumber \\
\hspace{-2.5cm}2\vec{{\bm \xi}}_3^{\mathrm{IX}} = - \sin \theta \sin \phi \partial_r + (\cos \phi - \cos \theta \cot r \sin \phi) \partial_{\theta} - (\cos \phi \cot r \csc \theta + \cot \theta \sin \phi) \partial_{\phi} , \nonumber \\
\end{eqnarray}
and the invariant fields are 
\begin{eqnarray}
\hspace{-2.5cm}2\vec{e}_1^{\mathrm{IX}} = -\cos \theta \partial_r + \cot r \sin \theta \partial_{\theta} - \partial_{\phi} \nonumber \\
\hspace{-2.5cm}2\vec{e}_2^{\mathrm{IX}} = - \cos \phi \sin \theta \partial_r + (\sin \phi - \cos \theta \cos \phi \cot r) \partial_{\theta} + (\cot r \csc \theta \sin \phi + \cos \phi \cot \theta) \partial_{\phi} \nonumber \\
\hspace{-2.5cm}2\vec{e}_3^{\mathrm{IX}} = - \sin \theta \sin \phi \partial_r - (\cos \phi + \cos \theta \cot r \sin \phi) \partial_{\theta} + (\cot \theta \sin \phi - \cos \phi \cot r \csc \theta ) \partial_{\phi} . \nonumber \\
\end{eqnarray}

\end{document}